\documentclass[a4paper,fleqn]{cas-sc}

\usepackage[numbers, sort&compress]{natbib}
\usepackage{amsmath}
\usepackage{amssymb}
\usepackage{caption}
\usepackage{graphicx}
\usepackage{latexsym}
\usepackage{nicefrac}
\usepackage{multirow}
\usepackage{multicol}
\usepackage{xcolor}
\usepackage{subcaption}
\usepackage{upgreek}
\usepackage{subfiles}

\usepackage{background}
\backgroundsetup{contents=Preprint, opacity=0.40, scale=15, color=gray, angle=45}

\def\tsc#1{\csdef{#1}{\textsc{\lowercase{#1}}\xspace}}
\tsc{WGM}
\tsc{QE}

\newcommand{\f}{\frac}
\newcommand{\sub}{\textsubscript}
\newcommand{\super}{\textsuperscript}
\newcommand{\nf}{\nicefrac}
\renewcommand{\d}{\mathrm{d}}
\newcommand{\dt}{\mathrm{d}t} 

\newcommand{\bbm}{\begin{bmatrix}}
\newcommand{\ebm}{\end{bmatrix}}

\newcommand{\beq}{\begin{equation}}
\newcommand{\eeq}{\end{equation}}
\newcommand{\beqnn}{\begin{equation*}}
\newcommand{\eeqnn}{\end{equation*}}
\newcommand{\bdis}{\begin{displaymath}}
\newcommand{\edis}{\end{displaymath}}
\newcommand{\beqarr}{\begin{eqnarray}}
\newcommand{\eeqarr}{\end{eqnarray}}
\newcommand{\beqarrnn}{\begin{eqnarray*}}
\newcommand{\eeqarrnn}{\end{eqnarray*}}

\renewcommand{\th}{\theta}
\newcommand{\lc}{L_\text{c}}
\newcommand{\tign}{T_\text{ign}}
\newcommand{\mfp}{\lambda_\text{MFP}}
\newcommand{\kg}{k_\text{g}}
\newcommand{\cbar}{\bar{c}}
\newcommand{\dpzero}{d_{\text{p},0}}
\newcommand{\hp}{H_\text{p}}
\newcommand{\rp}{r_\text{p}}

\newcommand{\ta}{T_\text{a}}

\newcommand{\ei}{E_\text{i}}
\newcommand{\er}{E_\text{r}}
\newcommand{\ep}{E_\text{p}}

\newcommand{\taone}{T_\text{a,1}}
\newcommand{\kinfone}{k_{\infty,1}}

\newcommand{\tai}{T_{\text{a},i}}
\newcommand{\mdotik}{\dot{m}_{i,\text{k}}}
\newcommand{\mdotid}{\dot{m}_{i,\text{d}}}

\newcommand{\NA}{\mathcal{N}_\mathrm{A}}
\newcommand{\RU}{\mathcal{R}_\text{u}}
\newcommand{\KB}{k_\text{B}}
\newcommand{\RG}{R_\text{g}}

\newcommand{\Kn}{\text{Kn}}
\newcommand{\Le}{\text{Le}}
\newcommand{\Dastar}{\text{Da\super{*}}}
\newcommand{\D}{\mathcal{D}}
\newcommand{\alphat}{\alpha_\text{T}}
\newcommand{\alphatj}{\alpha_{\text{T},j}}
\newcommand{\betat}{\beta_\text{T}}
\newcommand{\gt}{G_\text{T}}
\newcommand{\alpham}{\alpha_\text{M}}
\newcommand{\betam}{\beta_\text{M}}
\newcommand{\betaeff}{\beta_\text{eff}}
\newcommand{\gm}{G_\text{M}}

\newcommand{\fe}{\text{Fe}}
\newcommand{\rfe}{r_\fe}
\newcommand{\afe}{A_\fe}
\newcommand{\rhofe}{\rho_\fe}
\newcommand{\mfe}{m_\fe}
\newcommand{\hfe}{h_\fe}
\newcommand{\mdotfe}{\dot{m}_\fe}
\newcommand{\mdotfek}{\dot{m}_\text{Fe,k}}
\newcommand{\mdotfed}{\dot{m}_\text{Fe,d}}

\newcommand{\feo}{\text{FeO}}
\newcommand{\rfeo}{r_\feo}
\newcommand{\rhofeo}{\rho_\feo}
\newcommand{\xfeo}{X_\feo}

\newcommand{\mdotfeo}{\dot{m}_\feo}

\newcommand{\qfeo}{q_\feo}

\newcommand{\ftf}{\text{Fe\sub{3}O\sub{4}}}
\newcommand{\rftf}{r_\ftf}
\newcommand{\rhoftf}{\rho_\ftf}

\newcommand{\qftf}{q_\ftf}

\newcommand{\ftt}{\text{Fe\sub{2}O\sub{3}}}

\newcommand{\ox}{\text{O\sub{2}}}

\newcommand{\cox}{C_\ox}
\newcommand{\muox}{\mu_\ox}
\newcommand{\mdotox}{\dot{m}_\text{O\sub{2}}}
\newcommand{\mdotoxk}{\dot{m}_\text{O\sub{2},k}}
\newcommand{\mdotoxd}{\dot{m}_\text{O\sub{2},d}}

\newcommand{\inert}{\text{N\sub{2}}}

\newcommand{\cinert}{C_\inert}

\newcommand{\tg}{T_\text{g}}
\newcommand{\cg}{C_\text{g}}
\newcommand{\coxg}{C_\text{O\sub{2},g}}

\newcommand{\mg}{m_\text{g}}
\newcommand{\cbarg}{\bar{c}_\text{g}}
\newcommand{\wg}{W_\text{g}}
\newcommand{\fg}{f_\text{g}}
\newcommand{\gammag}{\gamma_\text{g}}

\newcommand{\mfpg}{\lambda_\text{MFP,g}}

\newcommand{\tth}{T_\theta}
\newcommand{\cth}{C_\theta}
\newcommand{\coxth}{C_{\text{O\sub{2}},\theta}}
\newcommand{\cinth}{C_{\text{N\sub{2}},\theta}}
\newcommand{\mth}{m_\theta}
\newcommand{\cbarth}{\bar{c}_\theta}

\newcommand{\qdotth}{\dot{Q}_\theta}
\newcommand{\mdotoxth}{\dot{m}_{\text{O\sub{2}},\theta}}

\newcommand{\tp}{T_\text{p}}
\newcommand{\tpmax}{T_\text{p,max}}
\newcommand{\tmelt}{T_\text{melt}}
\newcommand{\cp}{C_\text{p}}
\newcommand{\coxp}{C_\text{O\sub{2},p}}
\newcommand{\cinp}{C_\text{N\sub{2},p}}

\newcommand{\cbarp}{\bar{c}_\text{p}}

\newcommand{\qdotp}{\dot{Q}_\text{p}}
\newcommand{\mdotoxp}{\dot{m}_\text{O\sub{2},p}}

\newcommand{\qdot}{\dot{Q}}
\newcommand{\qdotr}{\dot{Q}_\text{R}}
\newcommand{\qdotc}{\dot{Q}_\text{C}}
\newcommand{\qdotfm}{\dot{Q}_\text{FM}}
\newcommand{\mdot}{\dot{m}}
\newcommand{\mdotc}{\dot{m}_\text{C}}
\newcommand{\mdotfm}{\dot{m}_\text{FM}}

\begin{document}
\let\WriteBookmarks\relax
\def\floatpagepagefraction{1}
\def\textpagefraction{.001}

%%%%%%%%%%%%%%%%%%%%%%%%%%%%%%%%%%%%%%%%%%%
% TITLE AND AUTHORS
%%%%%%%%%%%%%%%%%%%%%%%%%%%%%%%%%%%%%%%%%%%

\shorttitle{Knudsen regime iron particle ignition}    
\shortauthors{J. Jean-Philyppe et al.}  

\title [mode = title]{The ignition of fine iron particles in the Knudsen transition regime}  

\author[1]{Joel Jean-Philyppe}%[<options>]
\author[1]{Aki Fujinawa}
\author[1]{Jeffrey M. Bergthorson}
\author[2, 3]{XiaoCheng Mi}
\cormark[1]
\cortext[1]{Corresponding author\\
\indent\indent$^a$Department of Mechanical Engineering, McGill University, 845 Sherbrooke Street West, Montreal, QC, H3A 0G4, Canada\\
\indent\indent$^b$Power \& Flow Group, Department of Mechanical Engineering, Eindhoven University of Technology, 5600 MB, Eindhoven, the Netherlands\\
\indent\indent$^c$Eindhoven Institute of Renewable Energy Systems, Eindhoven University of Technology, 5600 MB, Eindhoven, the Netherlands}
\ead{x.c.mi@tue.nl}

%\affiliation[1]{organization={Department of Mechanical Engineering, McGill University},
%            addressline={845 Sherbrooke Street West}, 
%            city={Montreal},
%            state={QC},
%            postcode={H3A 0G4}, 
%            country={Canada}} 
%                       
%\affiliation[2]{organization={Power \& Flow Group, Department of Mechanical Engineering, Eindhoven University of Technology},
%            addressline={5600 MB}, 
%            city={Eindhoven},
%%            state={},
%%            postcode={}, 
%            country={the Netherlands}}
%            
%\affiliation[3]{organization={Eindhoven Institute of Renewable Energy Systems, Eindhoven University of Technology},
%            addressline={5600 MB}, 
%            city={Eindhoven},
%%            state={},
%%            postcode={}, 
%            country={the Netherlands}}

%%%%%%%%%%%%%%%%%%%%%%%%%%%%%%%%%%%%%%%%%%%
% ABSTRACT
%%%%%%%%%%%%%%%%%%%%%%%%%%%%%%%%%%%%%%%%%%%

\begin{abstract}
A theoretical model is considered to predict the minimum ambient gas temperature at which fine iron particles can undergo thermal runaway--the ignition temperature. The model accounts for Knudsen transition transport effects, which become significant when the particle size is comparable to, or smaller than, the molecular mean free path of the surrounding gas. Values of the thermal and mass accommodation coefficients for heat and mass transport are computed using a semi-empirical correlation. Two kinetic models for the high-temperature solid-phase oxidation of iron are analyzed. The first model (parabolic kinetics) considers the inhibiting effect of the iron oxide layers at the particle surface on the kinetic rate of oxidation, and a kinetic rate independent of the gaseous oxidizer concentration. The ignition temperature is solved as a function of particle size and initial oxide layer thickness with an unsteady analysis considering the growth of the oxide layers. In the free-molecular limit (small particles), the thermal insulating effect of transition heat transport can lead to a decrease of ignition temperature with decreasing particle size. However, the presence of the oxide layer slows the reaction kinetics and its increasing proportion in the small-particle limit can lead to an increase of ignition temperature with decreasing particle size. This effect is observed for sufficiently large initial oxide layer thicknesses. In that aspect, a steady analysis neglecting the growth of the oxide layer is shown to lead to an under-prediction of the ignition temperature, namely for large particles and small initial oxide layer thicknesses. The continuum transport model is shown to predict the ignition temperature of iron particles exceeding an initial diameter of 30~$\upmu$m to a difference of 3\% or less (30~K or less) when compared to the prediction of the transition transport model. In the small-particle limit and for sufficiently low oxidizer molar fractions in the bulk gas, the transition transport model reveals a combustion regime where the particle undergoes thermal runaway, followed by a stabilized combustion below the melting point of iron and its oxides. The second kinetic model (first-order kinetics) considers a porous, non-hindering oxide layer, and a linear  dependence of the kinetic rate of oxidation on the gaseous oxidizer concentration. The ignition temperature is resolved as a function of particle size with the transition and continuum transport models, and the differences between the ignition characteristics predicted by the two kinetic models are identified and discussed.
%In the free-molecular limit, the thermal insulating effect of transition transport, which facilitates ignition, competes with the rarefied oxidizer-particle collisions, which inhibit the reaction rate. The net result is an increase of ignition temperature with decreasing particle size. The small-particle ignition degeneration limit obtained with the continuum analysis is shown to be removed with the transition transport model, as the ignition temperature becomes independent on the particle size and plateaus for sufficiently small particles.
\end{abstract}

%%%%%%%%%%%%%%%%%%%%%%%%%%%%%%%%%%%%%%%%%%%
% KEYWORDS
%%%%%%%%%%%%%%%%%%%%%%%%%%%%%%%%%%%%%%%%%%%

\begin{keywords}
Iron particle \sep Metal fuel \sep Heterogeneous combustion \sep Ignition \sep Knudsen transition heat and mass transfer
\end{keywords}

%%%%%%%%%%%%%%%%%%%%%%%%%%%%%%%%%%%%%%%%%%%
% MAKE TITLE
%%%%%%%%%%%%%%%%%%%%%%%%%%%%%%%%%%%%%%%%%%%

\maketitle
\date{\today}

%%%%%%%%%%%%%%%%%%%%%%%%%%%%%%%%%%%%%%%%%%%
% BODY
%%%%%%%%%%%%%%%%%%%%%%%%%%%%%%%%%%%%%%%%%%%

\section{Introduction} \label{sec:intro}

Iron has an excellent potential as a global energy carrier due to its high energy density, its abundance, and the existing widely-developed iron mining, production, and recycling industries \cite{bergthorson2015, bergthorson2018}. To design and optimize practical iron burners, a deeper understanding of the physics underlying the combustion of fine iron particles is required in the scientific community. In particular, the ignition phenomenon of solid fuel particles leads to a burning regime exhibiting rapid reaction kinetics and high energy release rates \cite{frank1955, soo2018}. Iron burners with ignited particles therefore present the potential for practical, high-power applications, motivating the need to accurately predict iron particle ignition.

Conventionally, the continuum assumption has been adopted to describe transport processes in metal combustion problems. This assumption fails when the solid particles are of comparable size to, or smaller than, the gas molecular mean free path, as quantified by the Knudsen number (Kn), the ratio of the mean free path to the particle radius. Generally, researchers reported that for $\Kn \leq 0.01$, continuum treatment accurately describes transport processes; for $\Kn \geq 10$, free-molecular laws describe transport processes; and at intermediate Kn, transport occurs in the transition regime \cite{gopalakrishnan2011, kodas1990, liu2006, shpara2020, zou2020}.
%In a recent work, Ermoline \cite{ermoline2018} reported the Knudsen number alone may be misleading to assess heat transfer regimes. The ratio $(\Kn + \Kn^2)/(\alphat \kappa)$, which quantifies the contributions of free-molecular and continuum transport to transition heat transfer, was instead suggested, where $\alphat$ is the thermal accommodation coefficient (TAC), and $\kappa$ is a dimensionless parameter depending on the gas thermal conductivity and heat capacity ratio.
%%Ermoline's analysis did not consider transition effects on mass transfer.

In the past few decades, several studies have investigated the limits of applicability of continuum transport in heterogeneous reaction problems. In engineering systems involving the formation of aerosol nanoparticles and vapor molecules, Gopalakrishnan et al. \cite{gopalakrishnan2011} reported transition effects must be considered for submicron and nano- particles at 1 atm. Shpara et al. \cite{shpara2020} established the onset of transition effects between 1.23 and 46.3~$\upmu$m particle diameter for boron combustion between 4.0 and 0.1~MPa. In the heating and ignition delay time of metallic particles, Mohan et al. \cite{mohan2008} reported that these effects become important at 2 and 18~$\upmu$m particle diameter for pressures of 10 and 1~bar. Ermoline \cite{ermoline2018} reported transition heat transfer becomes significant for predicting the ignition of aluminum nano- and micro- particles at 1 atm. Recently, Senyurt and Dreizin \cite{senyurt2022} studied the ignition of aluminum, boron, and magnesium particles, and stated transition effects could be important up to 200~$\upmu$m particle diameter at 1~atm. 

Alas, transition effects were not previously captured in iron particle ignition problems. In a recent work, Mi et al. \cite{mi2022} investigated the ignition behavior of iron particles governed by a parabolic oxidation law. This kinetic model considers the hindrance of the transient oxide layer growth on the kinetic rate of solid-phase iron oxidation \cite{paidassi1958}, and an independence of the kinetic oxidation rate on the surrounding gaseous oxidizer concentration \cite{goursat1973}. The results were computed using a continuum transport model. In the current work, the analysis reported in Ref.~\cite{mi2022} is extended, by conducting a quantitative study of the ignition behavior of fine iron particles across the Knudsen transition regime, while applying the flux-matching boundary sphere method \cite{fuchs1959, liu2006} to resolve transition heat and mass transport. Additionally, different iron oxidation kinetics are investigated, namely the first-order kinetics proposed by Hazenberg and van Oijen \cite{hazenberg2021}. This kinetic model considers a porous iron oxide layer not hindering the kinetic rate of solid-phase iron oxidation, and a linear (first-order) dependence of the kinetic oxidation rate on the gaseous oxidizer concentration at the particle surface. 

The current work is structured as follows. In Section~\ref{sec:background}, an overview of available kinetic models for solid-phase iron oxidation is presented. Then, the mechanisms of gas-particle heat and mass transfer as a function of the Knudsen number are presented, and the physics underlying thermal and mass accommodation for transition and free-molecular transport are discussed. In Section~\ref{sec:model}, the model to predict ignition of a single iron particle accounting for transition heat and mass transport effects is formulated, with unsteady and steady analyses. In Section~\ref{sec:results}, the ignition temperature is resolved as a function of particle size for the parabolic and first-order kinetic models, and the results are compared to continuum transport modeling. Sources of error are discussed in Section~\ref{sec:discussion}, and concluding remarks are provided in Section~\ref{sec:conclusion}.

%\bibliographystyle{../pci}
%\bibliography{../afl_refs}

\section{Background} \label{sec:background}

\subsection{High-temperature solid-phase iron oxidation kinetics} \label{sec:kineticstheory}

Chen and Yeun \cite{chen2003} reviewed experimental data on the high-temperature solid-phase oxidation of iron (Fe) in air and oxygen. In the temperature range from 973--1523~K, the oxidation of Fe was found to consistently result in the formation of a three-layered, compact oxide scale adhering to the Fe surface, where the scale was composed of w\"{u}stite (FeO), magnetite (\ftf{}), and hematite (\ftt{}). As well, experimental evidence reported by Goursat and Smeltzer \cite{goursat1973} suggested the kinetic rate of oxidation to be independent of the surrounding gaseous oxidizer concentration, as the reaction rate-limiting step is the internal diffusion rate of iron ions through the oxide layers. In fact, the kinetics follow a parabolic rate law,
\beq \label{eq:parabolic}
X^2 = k_X t \Rightarrow \f{\d X}{\dt} = \f{k_X}{2 X}
\eeq
where $X$ is the thickness of the oxide layer, $k_X$ is the kinetic rate, and $t$ is time. Hence, the growth rate of the oxides were found to be inversely proportional to their increasing thickness. 

Following the parabolic kinetic model, Mi et al. \cite{mi2022} proposed an unsteady iron particle oxidation model considering the growth of a two-layered oxide scale on the particle surface. The iron oxides FeO and \ftf{} were formed through two parallel reactions between Fe and O, while the formation of \ftt{} was neglected, due to its relative negligible thickness in the oxide layer \cite{paidassi1958}. The growth rate of each oxide was formulated to follow Eq.~(\ref{eq:parabolic}), with kinetic parameters calibrated on the experimental work of Pa\"{i}dassi \cite{paidassi1958}. The independence of the kinetic rate on oxidizer concentration led to a specialization of the $k$-$\beta$ model \cite{soo2018} to a switch-type model, which naturally captures the transition between kinetic- and external-diffusion- limited combustion. 

An alternative kinetic model was recently proposed by Hazenberg and van Oijen \cite{hazenberg2021} to model the steady one-dimensional propagation of flames in iron particle suspensions. They considered spherical particles composed of a Fe core surrounded by a growing porous FeO shell which does not inhibit the kinetic rate of oxidation. Instead, a first-order (linear) dependence of the kinetic reaction rate on the oxidizer concentration at the particle surface was assumed. In a different study by Lysenko et al. \cite{lysenko2014}, the oxidation kinetics of iron particles were experimentally studied through thermogravimetric analysis (TGA) in the temperature range from 298--1073~K. The oxidation kinetics were formulated as a set of $n$-order branched reactions, where two parallel pathways were considered: direct formation of \ftt{} from Fe and O; and formation of \ftf{} from Fe and O, then further oxidation of \ftf{} to \ftt{}. 

Among the three kinetic models presented, the model described by Mi et al. \cite{mi2022} is proposed to better represent the underlying physics of high-temperature solid-phase iron oxidation, as it is based on an extensive experimental and theoretical literature on the topic. In the model proposed by Lysenko et al. \cite{lysenko2014}, the temperature range covered (298--1073~K) maps multiple different oxidation mechanisms, as reviewed by Chen and Yeun \cite{chen2003}, namely since FeO becomes a stable oxide above 843~K. However, this was not captured by Lysenko et al. \cite{lysenko2014}, who proposed a single set of equations covering this entire temperature range. Regarding the model proposed by Hazenberg and van Oijen \cite{hazenberg2021}, literature shows iron oxidation results in a multi-layered oxide scale dependent on temperature \cite{chen2003}, beyond the formation of only FeO. As well, FeO was assumed to be porous, despite its Pilling-Bedworth ratio of 1.7 predicts a compact, protective oxide layer \cite{xu2000}. Nevertheless, the current work implements both the kinetic models proposed in Ref.~\cite{mi2022} and \cite{hazenberg2021}: the former implementation permits a characterization of transition transport effects on parabolic kinetics independent of oxidizer concentration, a realistic representation of iron oxidation kinetics, while the latter allows an elucidation of the fundamental impact of Knudsen transition effects on a first-order kinetic reaction.

%\bibliographystyle{../pci}
%\bibliography{../afl_refs}

\subsection{Gas-particle heat and mass transfer} \label{sec:heatmasstransfer}

%%%%%%%%%%%%%%%%%%%%%%%%%%%%%%%%%%%%%%%%%%%
% TRANSPORT REGIMES
%%%%%%%%%%%%%%%%%%%%%%%%%%%%%%%%%%%%%%%%%%%

\subsubsection{Knudsen number and transport regimes}

Intrinsic to the solid-phase iron oxidation and ignition problem is heterogeneous heat and mass transport. It occurs under different regimes as a function of the non-dimensional Knudsen number, $\Kn = \mfp / \lc$, the ratio of the gas Maxwell molecular mean free path, $\mfp$, to the system characteristic length, $\lc$. In the case of a particle, the characteristic length is typically defined as the particle radius, $\lc = \rp$. When the Knudsen number is much greater than 1 (e.g. $\Kn \geq 10$), transport processes occur in the \textit{free-molecular regime} \cite{gopalakrishnan2011, kodas1990, liu2006, shpara2020, zou2020}. The gas molecules on average travel a large distance between inter-molecular collisions, and a low molecule-particle collision rate is observed, which limits heat transfer, and controls mass transfer in external-diffusion-limited heterogeneous combustion. In the free-molecular regime, the analytical solution to the rates of heat and mass transport for a stationary particle in a quiescient gas yields \cite{kennard1938, liu2006, qu2001},
\beqarr
\label{eq:qdotfm} \qdotfm &=& \alphat \pi \rp^2 \f{p \cbarg}{2} \bigg(\f{\gamma+1}{\gamma-1}\bigg) \bigg(\f{\tp}{\tg} - 1\bigg) \\
\label{eq:mdotfm} \mdotfm &=& \alpham \pi \rp^2 (\cg \cbarg - \cp \cbarp)
\eeqarr
where $\alphat$ is the thermal accommodation coefficient (TAC), $p$ is the bulk gas pressure, $\cbar$ is the gas average molecular speed, $\gamma$ is the gas heat capacity ratio, $T$ is temperature, $\alpham$ is the mass accommodation coefficient (MAC), and $C$ is the concentration of the gas species. The subscripts "p" and "g" indicate that parameters are evaluated at the particle surface and in the bulk gas, respectively. 

When the molecular mean free path is much smaller than the particle radius (e.g. $\Kn \leq 0.01$), transport processes occur in the \textit{continuum regime} \cite{gopalakrishnan2011, kodas1990, liu2006, shpara2020, zou2020}. Continuum transport is characterized by large inter-molecular and molecule-particle collision rates. The rate of heat transfer is limited by the inability of the molecules colliding with the particle to effectively carry the energy away from the surface to the bulk gas before encountering several collisions, while mass diffusion is also limited by the inter-molecular collisions. In the continuum regime, the analytical solutions to Fourier's law of conduction and Fick's law of diffusion yield,
\beqarr
\label{eq:qdotc} \qdotc &=& 4 \pi \rp k (\tp - \tg) \\
\label{eq:mdotc} \mdotc &=& 4 \pi \rp \D (\cg - \cp)
\eeqarr
where $k$ is the gas mixture-averaged thermal conductivity, and $\D$ is the mass diffusivity of the gas species in the mixture.

\begin{figure}
    \centering
    \includegraphics[width = 0.6\linewidth]{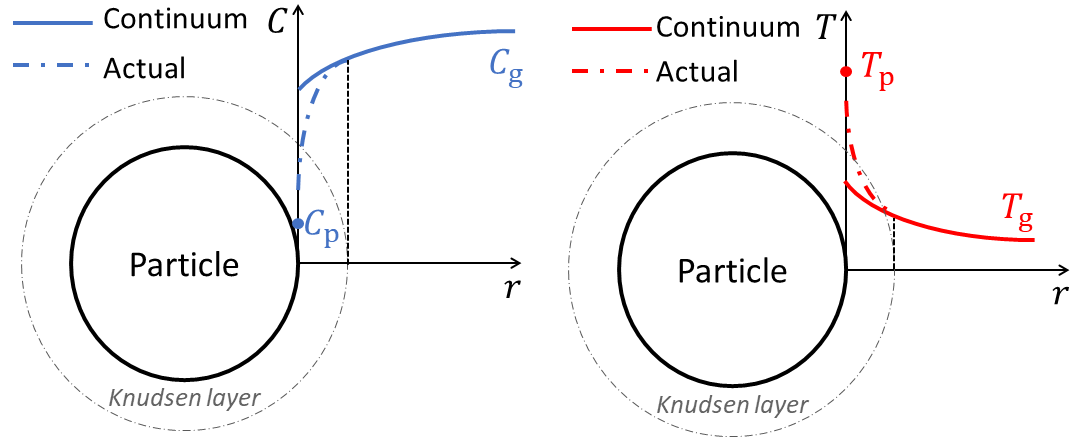}
    \caption{Temperature and concentration jump in the Knudsen layer. The profiles differ from the continuum solutions from the particle surface to the Knudsen layer boundary, as a result of gas kinetics effects on heterogeneous heat and mass transfer.}
    \label{fig:jump}
\end{figure}

For intermediate Kn (e.g. $0.01 < \Kn < 10$), transport processes occur in the \textit{transition regime} \cite{gopalakrishnan2011, kodas1990, liu2006, shpara2020, zou2020}. Conceptually speaking, the transition regime can be understood by delimiting two distinct regions around the particle surface \cite{kodas1990, liu2006}. In the vicinity of the particle, there exists a region with few inter-molecular collisions called the Knudsen layer, where transport mechanisms are governed by free-molecular gas kinetics. The thickness of the Knudsen layer is on the order of the gas molecular mean free path. Beyond the Knudsen layer, the transport mechanisms occur in continuum, where macroscopic heat conduction and mass diffusion laws apply. This conceptual understanding is based on experimental evidence of the temperature and concentration jumps arising from transport at heterogeneous interfaces, causing a deviation from the temperature and concentration profiles predicted by continuum, as depicted in Fig.~\ref{fig:jump} \cite{fuchs1959, goodman1976, kennard1938, liu2006, wagner1982, wright1960}. The Knudsen layer also exists in the continuum and free-molecular limits. However, its small size relative to the particle in the continuum limit leads to a negligible impact of gas kinetics transport effects at the particle surface, permitting solutions described purely by the macroscopic transport equations. Conversely, its large size relative to the particle in the free-molecular limit implies that continuum transport beyond the Knudsen layer need not be considered. 

The transition regime presents an additional difficulty, as there exists no analytical closed-form solution to describe the transport mechanisms in this range of Kn. Heat and mass transfer are governed by the full Boltzmann equation, an integral-differential equation with initial and boundary conditions \cite{kodas1990, liu2006}. Consequently, multiple modeling approaches to describe the transition regime were developed over the past decades. An elaborate review of transition heat conduction modeling employed in the laser-induced incandescence literature is presented by Liu et al. \cite{liu2006}, whereas several methods for heat and mass transport prediction in droplet evaporation problems are reported by Wagner \cite{wagner1982}. In the current work, the boundary sphere transition transport method--also called Fuchs' method--is employed.

%%%%%%%%%%%%%%%%%%%%%%%%%%%%%%%%%%%%%%%%%%%
% BOUNDARY SPHERE METHOD
%%%%%%%%%%%%%%%%%%%%%%%%%%%%%%%%%%%%%%%%%%%

\subsubsection{Modeling the transition regime - the boundary sphere method}

The boundary sphere flux-matching method consists in explicitly introducing a Knudsen layer of thickness $\th$ closely related to the mean free path of the gas molecules in the vicinity of the particle, delimiting the free-molecular and continuum transport regions. The sphere of radius $\rp + \th$ is called the limiting sphere, and its surface has temperature $\tth$ and gas concentration $\cth$. Immediately at the particle surface, the temperature and concentration are $\tp$ and $\cp$, while in the bulk gas, they are $\tg$ and $\cg$. The gradients between the particle and limiting sphere surfaces give rise to heat and mass transport described by free-molecular laws, and the gradients beyond the limiting sphere result in heat and mass transport described by continuum laws. The Knudsen layer is assumed to be in quasi-steady equilibrium, hence energy and mass conservation provide necessary conditions for flux-matching of the transport rates at the particle and limiting sphere surfaces. The boundary sphere method solves for the thickness of the Knudsen layer $\th$, as well as the temperature and species concentration at the jump distance which satisfy the conservation laws, $\tth$ and $\cth$, respectively. In the case of the combustion of a particle, two additional equations are needed to solve for $\tp$ and $\cp$: one for the heat generation rate from the particle, and one for the consumption rate of oxidizer. 

The mathematical formulation for the boundary sphere method is presented in Section~\ref{sec:balanceknudsen}. The boundary sphere method imposes no restrictions on the temperature and concentration differences between the particle surface and the bulk gas, while several other methods for example implicitly assume small temperature difference \cite{liu2006}. As well, this method yields the correct solutions in the free-molecular and continuum limits. 

%\bibliographystyle{../pci}
%\bibliography{../afl_refs}

\subsection{Accommodation coefficients} \label{sec:accommodation}

The TAC and MAC are critical in determining the transport rates in the free-molecular and transition regimes, and their quantitative determination are complicated problems which have been extensively studied in the literature over the past several decades. Nevertheless, there remains large uncertainty in determining their values for transport calculations between different solid-gas pairs, which is a challenge for iron particle ignition calculations. The determination of accommodation coefficients is rooted in the theory of gas-surface scattering processes, and a comprehensive review of this topic is beyond the scope of the current work. Instead, some selected pioneering studies on the TAC and MAC are mentioned, and practical considerations for their quantification in engineering applications are highlighted.

%%%%%%%%%%%%%%%%%%%%%%%%%%%%%%%%%%%%%%%%%%%
% DEFINITION
%%%%%%%%%%%%%%%%%%%%%%%%%%%%%%%%%%%%%%%%%%%

\subsubsection{Definition of accommodation coefficients}

The formal definition of the TAC, also referred to as the energy accommodation coefficient, is attributed to Knudsen and yields \cite{goodman1976, kennard1938, saxena1989},
\beq \label{eq:tacdef}
\alphat = \f{\langle \er - \ei \rangle}{\langle \ep - \ei \rangle}
\eeq
where $\ei$ and $\er$ are the energy of the incident and scattered gas molecules, $\ep$ is the energy the scattered gas molecules would carry if they reached thermal equilibrium with the wall before being scattered, and $\langle \cdot \rangle$ indicates an average. The denominator can be reformulated as $\langle \er - \ei \rangle_\text{max}$; hence, the TAC is a quantification of the effectiveness of heat transfer between the surface and the gas molecules upon molecule-surface encounters. In Eq.~(\ref{eq:tacdef}), energy is often equivalently substituted by temperature. Analogously, the MAC, also referred to as sticking, trapping, or adsorption coefficient, is defined as \cite{king1978, kodas1990}:
\beq \label{eq:macdef}
\alpham = \f{\text{net rate of adsorption}}{\text{rate of collisions}}.
\eeq
Hence, the MAC is a measure of the effectiveness of heterogeneous mass transfer upon molecule-surface encounters. In Eq.~(\ref{eq:macdef}), the rate of collisions can be determined from gas kinetics theory, while the net rate of adsorption takes into account both the rates of adsorption and desorption of gas molecules from the surface. The determination of the MAC is therefore based on the kinetics of surface processes. 

%%%%%%%%%%%%%%%%%%%%%%%%%%%%%%%%%%%%%%%%%%%
% GAS-SURFACE SCATTERING
%%%%%%%%%%%%%%%%%%%%%%%%%%%%%%%%%%%%%%%%%%%

\subsubsection{Gas-surface scattering processes and physical mechanisms of accommodation}

\begin{figure}[h]
	\centering
	\begin{subfigure}{0.35\linewidth}
	\centering
	\includegraphics[width=\linewidth]{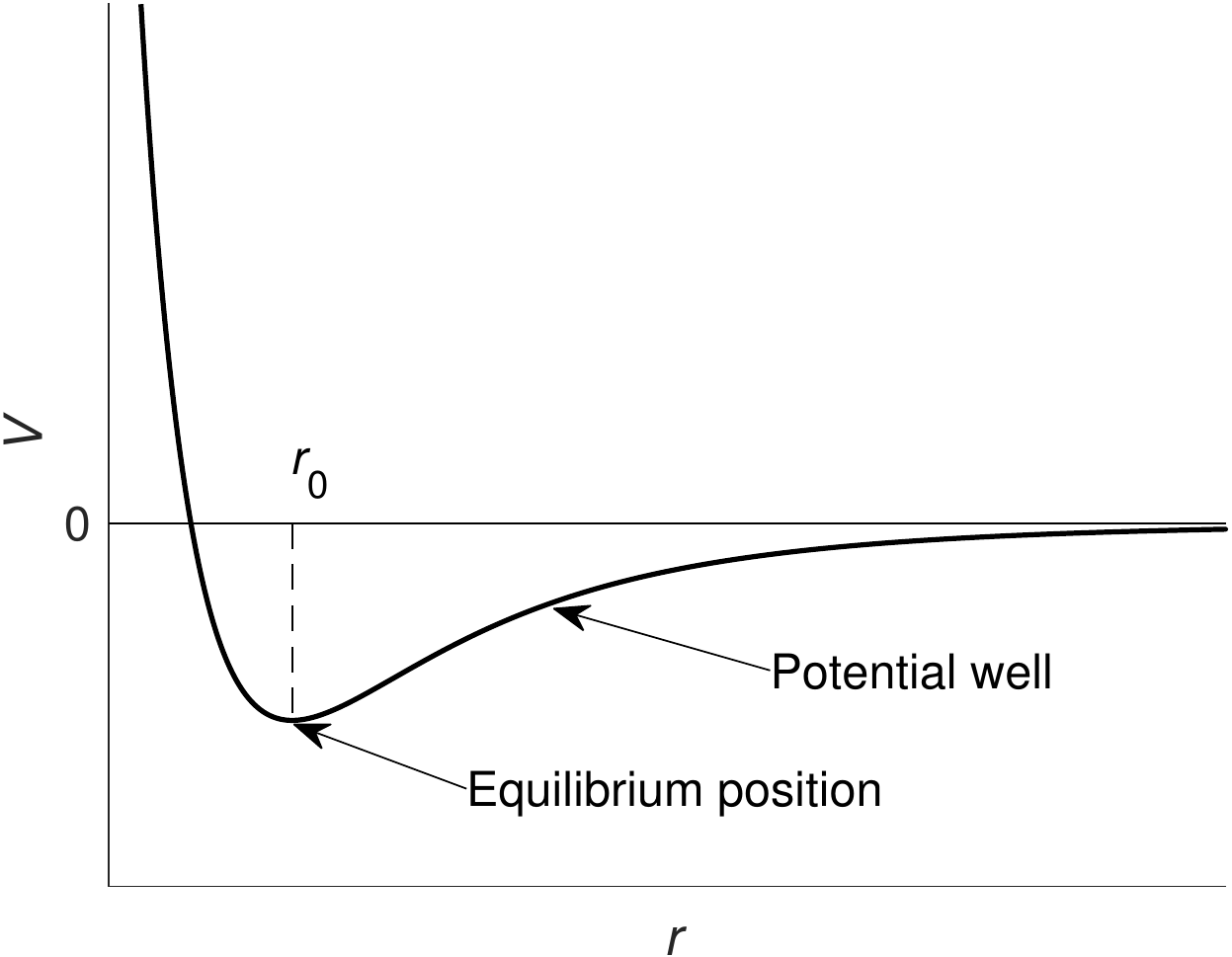}
	\caption{}
	\label{fig:potentialgeneric} 
	\end{subfigure} 
	\begin{subfigure}{0.35\linewidth}
	\centering
	\includegraphics[width=\linewidth]{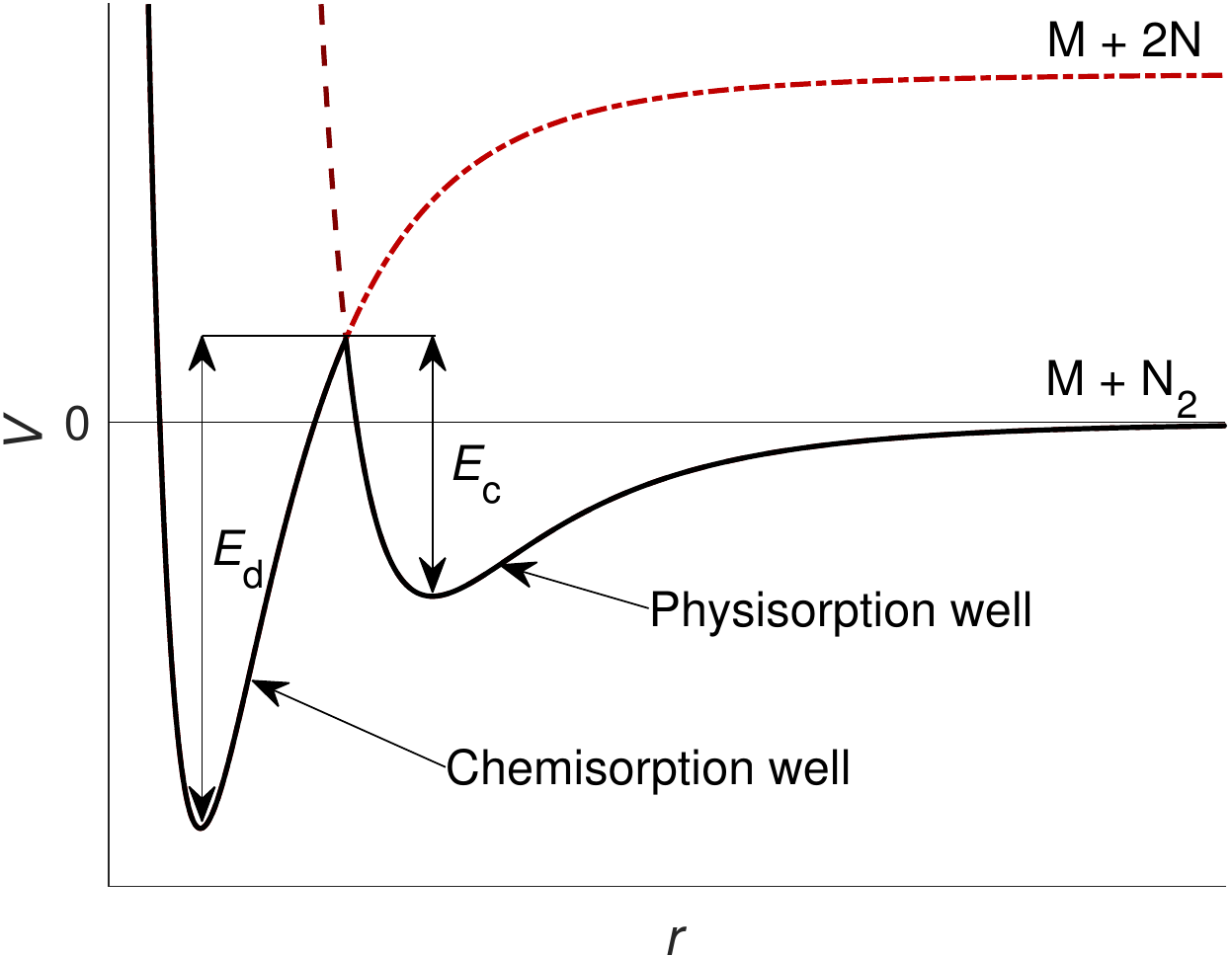}
	\caption{}
	\label{fig:potentialoverlap} 
	\end{subfigure}
	\caption{Schematic of interatomic potential $V$ as a function of interatomic distance $r$ for a solid and a gas atom. (a) The region $V(r) <  0$ is the potential well and gives rise to attractive forces, while $V(r) > 0$ results in repulsive forces. The minimum of the curve $r_0$ is the equilibrium position. The difference $[\lim_{r \to \infty} V(r)] - V(r_0)$ is the potential well depth. (b) Example of an overlap of the physisorption and chemisorption potential functions between a solid atom M and nitrogen. The physisorption potential relates to non-dissociative adsorption of N\sub{2}, while the chemisorption potential pertains to dissociative adsorption of individual N atoms. The molecule must overcome an activation barrier $E_\text{c}$ to enter the chemisorbed state, while it must overcome an activation barrier $E_\text{d}$ to desorb from the chemisorbed state.}
	\label{fig:potential} 
\end{figure}

The theory of gas-surface scattering processes is fundamentally based on the interaction of atoms through interatomic potential force fields \cite{barker1984, goodman1976}. From elementary gas kinetics theory, it is generally accepted that at large distances, physical van der Waals interactions give rise to attractive forces between atoms, while at short distances, the overlapping of electronic clouds results in repulsive forces. This is best demonstrated through a potential energy curve, as schematically depicted in Fig.~\ref{fig:potentialgeneric}. Upon approaching a solid surface, there exists various processes through which a gas atom or molecule may be scattered \cite{barker1984, goodman1976}. The interaction dynamics may lead to elastic or inelastic scattering, the latter implying energy transfer or accommodation through the TAC.

A number of key factors which govern the TAC were discussed by Goodman and Wachman \cite{goodman1976}, such as: the molecular mass ratio between the gas and solid, $\phi = m_\text{g} / m_\text{s}$; the surface roughness; and the adsorption state of gas atoms on the surface. A large ratio $\phi$ implies multiple interatomic collisions are required to reverse the momentum of the incident gas, which leads to a longer residence time near the solid and a higher probability of energy exchange through phonon or electronic processes. Additionally, large values of $\phi$ are typically associated with deeper interatomic interaction potentials. An increase of surface roughness leads to an increase in the TAC, when the roughness is on a scale which may cause multiple collisions of incident gas atoms or molecules before they are scattered back to the bulk gas. Moreover, the TAC increases with the presence of adsorbate gas layers on the surface when compared to the clean-surface TAC, due to enhanced energy transfer through gas-gas interactions, which are inherently stronger in nature than gas-solid interactions \cite{shin1965}. The surface temperature also has an indirect impact on the TAC: An increase in surface temperature induces a thermal roughening effect, which increases the clean-surface TAC \cite{sipkens2018}. However, increasing temperature induces desorption of the gas at the solid surface, which decreases the TAC. Experimental evidence has shown the latter effect outweighs the former, and the net result is a decrease in the TAC with increasing surface temperature \cite{goodman1976, song1987}.

Inelastic scattering, which gives rise to thermal accommodation, may occur through a direct pathway which involves phonon energy exchange between the gas and solid at short distances, or through a trapping-desorption (adsorption-desorption) process \cite{barker1984}, which is intrinsically linked to mass accommodation. 
%By applying the theory of interatomic potentials, Glasstone et al. \cite{glasstone1941} addressed the topic of adsorption and desorption since the early 1940s in their pioneering work on the kinetic theory of rate processes. They formulated adsorption and desorption as activated processes, and they derived kinetic expressions to calculate their rates from first principles, which based on Eq.~(\ref{eq:macdef}) would permit a determination of the MAC. 
King \cite{king1978} reviewed the kinetics of adsorption and desorption processes, while providing insight on their relationship with thermal and mass accommodation. The author noted that, upon approaching the solid surface, a gas molecule may be elastically scattered back to the gas phase--without energy exchange (no accommodation). Conversely, the incident gas molecule may lose sufficient translational kinetic energy to the surface to become adsorbed in the physisorbed state, i.e. being trapped in the van der Waals attractive potential well. For trapping to occur, the gas molecule must be completely thermally accommodated by the surface. Then, the molecule may be inelastically scattered back to the bulk gas (desorption); or if the formation of a chemisorbed state is possible between the solid-gas pair, it may pass to this more stable state after overcoming an activation barrier. The overlap of the physisorption and chemisorption potential wells is depicted in Fig.~\ref{fig:potentialoverlap}. Based on the physical definition of the MAC, and the fact that to remain trapped in an adsorbed state, the incident gas molecule must be completely thermally accommodated, King \cite{king1978} noted that there is an intrinsic relationship between the TAC and the MAC, which he inferred from a review of experimental measurements to be nearly one-to-one. 
%This interpretation is namely valid for gas-surface scattering between a solid-gas pair which is dominated by trapping-desorption processes, which is the case for deep interatomic interaction potentials.

%%%%%%%%%%%%%%%%%%%%%%%%%%%%%%%%%%%%%%%%%%%
% QUANTITATIVE
%%%%%%%%%%%%%%%%%%%%%%%%%%%%%%%%%%%%%%%%%%%

\subsubsection{Determination of accommodation coefficients for engineering surfaces}

A number of classical theories based on the dynamics of gas-surface scattering processes have been developed to calculate the TAC and MAC from first principles \cite{goodman1976, glasstone1941, king1978, saxena1989}. However, several assumptions, or the necessity to know parameters which are difficult to estimate, limit the applicability of these theories in practical calculations. Additionally, methods for experimental determination of the TAC and MAC have been extensively developed in the literature \cite{king1978, saxena1989}, and semi-empirical correlations typically provide good agreement with experimental data. However, the availability of experimental data is limited to specific solid-gas pairs and surface conditions, and their extrapolation is not straightforward. Furthermore, the treatment of the TAC and MAC has often been decoupled in the literature, despite their intrinsic relationship (e.g. in trapping-desorption-dominated inelastic scattering). An additional limitation is that the vast majority of experimental measurements of the TAC are carried out under clean-surface conditions, where impurities--adsorbate gas layers--have been thoroughly cleaned from the sample. The cleaning method involves bringing the surface to a very high temperature to induce desorption of all gas impurities, and cooling in a vacuum-controlled environment before introducing the test gas \cite{goodman1976, song1987}. Hence, although the clean-surface TAC provides important insight on the underlying physical processes which govern it, such experimental data is of limited relevance in practical engineering calculations. 

In an effort to address this issue, Song and Yovanovich \cite{song1987} proposed in the late 1980s a generic semi-empirical correlation based on the classical theory of the modified Baule formula \cite{goodman1976}. The formulation accounts for adsorbed gas layers through the surface temperature, and accounts for the molecular mass ratio $\phi$. The correlation was calibrated to experimentally-measured values of the TAC where no efforts were undertaken to clean the surface, and it is inferred to provide a good approximation of the TAC for engineering surfaces. In the current work, the Song and Yovanovich \cite{song1987} correlation is used to approximate the TAC, and the MAC of oxygen on iron is assumed to be equal to the TAC calculated through the correlation. This assumption is justified by the deep interaction potential between iron and oxygen, and it is inferred that trapping-desorption dominates accommodation between this solid-gas pair. 

%\bibliographystyle{../pci}
%\bibliography{../afl_refs}

%\subfile{sections/2.1-kinetics.tex}
%\subfile{sections/2.2-heat_mass_transfer.tex}
%\subfile{sections/2.3-accommodation.tex}
%\bibliographystyle{../pci}
%\bibliography{../afl_refs}

%\subfile{sections/2.1-kinetics.tex}
%\subfile{sections/2.2-heat_mass_transfer.tex}
%\subfile{sections/2.3-accommodation.tex}

\section{Model formulation for an isolated iron particle} \label{sec:model}

%%%%%%%%%%%%%%%%%%%%%%%%%%%%%%%%%%%%%%%%%%%
% MODEL DESCRIPTION
%%%%%%%%%%%%%%%%%%%%%%%%%%%%%%%%%%%%%%%%%%%

\subsection{Model description}

\begin{figure}[h]
    \centering
    \includegraphics[width = 0.4\linewidth]{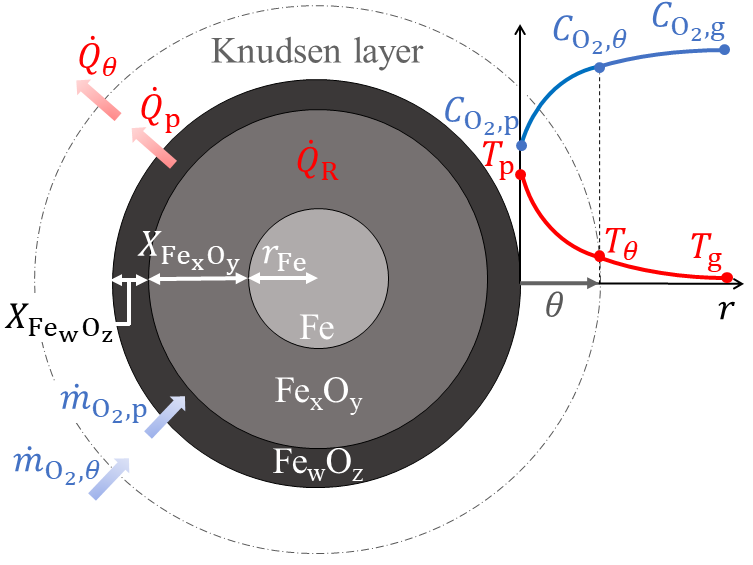}
    \caption{Particle reaction model. Symbols $T$ and $\cox$ denote temperature and oxidizer concentration. Subscripts "p", "$\th$", and "g" denote particle surface, Knudsen layer surface, and bulk gas. Heat and mass transport, $\qdot$ and $\mdotox$, are governed by free-molecular and continuum laws within and beyond the Knudsen layer of thickness $\th$, respectively.}
    \label{fig:model}
\end{figure}

\begin{figure}
    \centering
    \includegraphics[width = 0.5\linewidth]{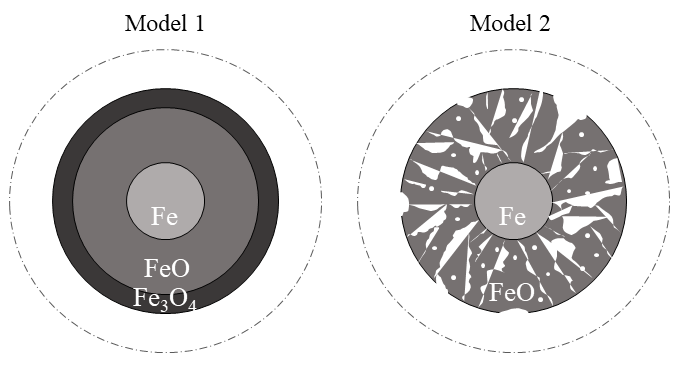}
    \caption{Solid-phase iron oxidation models. Model 1: multi-layered parabolic kinetic model with the formation of protective \feo{} and \ftf{} shells. Model 2: first-order reaction with the formation of a porous \feo{} shell.}
    \label{fig:oxidation}
\end{figure}

The current work couples solid-phase iron oxidation kinetics with the boundary sphere  flux-matching method to predict single iron particle ignition accounting for Knudsen transition heat and mass transport effects. The model considers a spherical particle consisting of an iron core of radius $\rfe$, surrounded by concentric iron oxide layers of thickness $X_i$, where $i$ represents the solid-phase oxides, and a Knudsen layer of thickness $\th$, as illustrated in Fig.~\ref{fig:model}. Two particle reaction models are considered, as illustrated in Fig.~\ref{fig:oxidation}:
\begin{enumerate}
\item a parabolic kinetic model with two parallel single-step reactions, based on the model of Mi et al. \cite{mi2022}, wherein $i$ takes the values "FeO" and "\ftf{}";
\item a first-order single-step kinetic model, based on the model of Hazenberg and van Oijen \cite{hazenberg2021}, wherein $i$ takes the value "FeO".
\end{enumerate}
As mentioned in Section~\ref{sec:kineticstheory}, the high-temperature parabolic oxidation of iron results  in the formation of a three-layered oxide scale on the iron surface, where \ftt{} is the outermost oxide layer. However, since its thickness is only 1\% of the total thickness of the oxides, its contribution to heat release in the particle is negligible, hence, it is neglected in the thermophysical analysis \cite{mi2022}.

The Knudsen layer thickness $\th$ is on the order of the gas molecular mean free path, and the sphere of radius $\rp + \th$ is the limiting sphere, where $\rp$ is the particle radius. In this region, there are few inter-molecular collisions, hence free-molecular laws describe transport processes, while macroscopic continuum laws describe heat and mass transport beyond the limiting sphere \cite{liu2006}. The particle is placed in hot air represented by a binary gas mixture consisting of 21\% oxygen ($\ox$) and 79\% nitrogen ($\inert$) by volume. Heat transfer between the particle and the gas occurs through conduction; radiation is neglected, since it has a negligible contribution to heat transfer in the pre-ignition phase \cite{mi2022}. Other assumptions of the model include:
\begin{enumerate}
    \item The solids maintain a constant density throughout the high-temperature oxidation and ignition process.
    \item Since the Biot number is small for the particle sizes considered, the particle is at a uniform temperature $\tp$.
    \item The bulk gas is in sufficient quantity such that its temperature $\tg$ and composition are not affected by the single particle ignition.
    \item The gas flow velocity is negligible.
    \item The Stefan flow is neglected.
    \item The second-order heat and mass transport mechanisms (Dufour and Soret effects) are neglected.
\end{enumerate}

The particle internal energy (or enthalpy), $\hp$, is tracked in time, along with its mass content in each solid-phase species, $\mfe$ and $m_i$, where the oxides $i$ are determined by the particle kinetics. The enthalpy is formulated as,
\beq \label{eq:hp}
\hp = \mfe \hfe(\tp) + \sum_i m_i h_i(\tp)
\eeq
where $\hfe$ and $h_i$ represent the specific gravimetric enthalpy of the solids, computed as a function of $\tp$ with the Shomate equation based on the NIST Database \cite{chase1991}. While $\hp$ is tracked, an iterative root-finding procedure can be applied to resolve $\tp$.

%%%%%%%%%%%%%%%%%%%%%%%%%%%%%%%%%%%%%%%%%%%
% CONSERVATION LAWS
%%%%%%%%%%%%%%%%%%%%%%%%%%%%%%%%%%%%%%%%%%%

\subsection{Conservation laws in the Knudsen layer} \label{sec:balanceknudsen}

\subsubsection{Implicit method}

The boundary sphere heat balance yields $\qdotp = \qdotth \equiv \qdot$, where $\qdotp$ is the free-molecular heat loss rate from the particle surface to the Knudsen layer, and $\qdotth$ is the continuum heat loss rate from the limiting sphere surface to the bulk gas. Similarly, the mass balance yields $\mdotoxp = \mdotoxth \equiv \mdotox$, where $\mdotoxp$ is the free-molecular \ox{} consumption rate at the particle surface, and $\mdotoxth$ is the continuum \ox{} diffusion rate at the limiting sphere surface. Using Eq.~(\ref{eq:qdotfm})--(\ref{eq:mdotc}) yields:
\beqarr
\label{eq:bsheat} \alphat \pi \rp^2 \f{p \cbarth}{2} \bigg(\f{\gamma^*+1}{\gamma^*-1} \bigg) \bigg(\f{\tp}{\tth} - 1\bigg) &=& 4 \pi (\rp + \th) k^* (\tth - \tg) \\
\label{eq:bsmass} \alpham \pi \rp^2 (\coxth \cbarth - \coxp \cbarp) &=& 4 \pi (\rp + \th) \D^* (\coxg - \coxth).
\eeqarr

Since the Stefan flow is neglected and the consumption rate of oxidizer is small in the pre-ignition phase, the pressure $p = \KB \NA T (\sum_j C_j / W_j)$ is spatially and temporally uniform, where $\KB$ is the Boltzmann constant, $\NA$ is Avogadro's number, $W_j$ is the molar weight of the gaseous species, and $j$ takes the values "\ox{}" and "\inert{}". With this formulation, given $\cox$ and $T$ at a given location, "p" or "$\th$", the corresponding $\cinert$ can be computed. The gas average molecular speed is $\cbar = [8 \KB \NA T/(\pi W)]^{1/2}$, where $W = (\sum_j C_j W_j) / (\sum_j C_j)$ is the gas mixture-averaged molar weight. The thickness of the Knudsen layer is formulated as the mean free path evaluated in the bulk gas, $\th = \mfpg$, computed as \cite{liu2006},
\beq \label{eq:mfpg}
\mfpg = \f{ \pi \kg (\gammag-1) \wg \cbarg }{4 \KB \NA \fg p }
\eeq
where $\fg = (9 \gammag - 5)/4$ is the Eucken factor. Since the bulk gas properties remain constant throughout the ignition process, $\th$ is constant.

In Eq.~(\ref{eq:bsheat}) and (\ref{eq:bsmass}), the superscript \super{*} indicates that the mixture-averaged thermophysical and transport properties are evaluated with a two-third law \cite{hubbard1975}, wherein $\gamma^*$ is evaluated closer to $\{\tp, \, \coxp, \, \cinp\}$ than the corresponding properties at $\th$, while $\{k^*, \, \D^*\}$ are evaluated closer to $\{\tth, \, \coxth, \, \cinth\}$ than their counterpart in the bulk gas. The heat capacity ratio is approximated as $\gamma = c_p / [c_p - \RU / W]$, where $c_p$ is the mixture gravimetric heat capacity at constant pressure, $c_p = (\sum_j C_j c_{p,j}) / (\sum_j C_j)$, and $\RU$ is the universal gas constant. The heat capacities of the species $c_{p,j}$ are evaluated as a function of $T$ with the NASA 7-coefficients polynomials \cite{mcbride1993}. The mixture thermal conductivity is approximated with the Wilke mixture rule, 
\beq
k = \f{1}{2} \bigg(\sum_j \mu_j k_j + \f{1}{\sum_j (\mu_j/k_j) } \bigg)
\eeq
where $\mu_j$ is the molar fraction of the gas species, and $k_j$ is its thermal conductivity, computed as a function of $T$ with the NASA 5-coefficients polynomials \cite{mcbride1993}. The oxidizer mass diffusivity in the binary mixture is computed with the Fuller-Schettler-Giddins semi-empirical correlation \cite{fuller1966}:
\beq
\D = \f{10^{-7} T^{7/4} \big[ \sum_j (1/W_j) \big] ^{1/2} }
{ \f{p}{101325} \big[ \sum_j \big(v_j^{1/3}\big) \big] ^ 2 }.
\eeq
Values for the semi-empirical parameters $v_j$ can be found in Table~\ref{tab:properties}.

Equations~(\ref{eq:bsheat}) and (\ref{eq:bsmass}) require a method to approximate the TAC and MAC, $\alphat$ and $\alpham$. For each species in the gas mixture, the TAC is computed with the Song and Yovanovich \cite{song1987} semi-empirical correlation:
\beq \label{eq:tacsong}
\alphatj = F \bigg( \f{W_j}{6.80 + W_j} \bigg) + (1-F) \bigg(  \f{2.40 \NA \phi_j}{(1+\phi_j)^2} \bigg)
\eeq
where $F = \exp[-0.57(\tp-273)/273]$, $\phi_j = W_j/W_\text{oxide}$, and $W_\text{oxide}$ is the molar weight of the oxide at the external surface of the particle--\ftt{} or FeO. Although \ftt{} is neglected in the thermophysical analysis of the parabolic kinetic model, it is used to compute $\alphatj$, which is a surface property. The mixture TAC is computed as \cite{mikami1966}:
\beq
\alphat = \bigg( \sum_j \f{\mu_j \alphatj}{W_j^{\nf{1}{2}}}\bigg) \bigg/ \bigg( \sum_j \f{\mu_j}{W_j^{\nf{1}{2}}} \bigg).
\eeq
The MAC of \ox{} on the surface is set equal to the corresponding TAC of \ox{} computed with Eq.~(\ref{eq:tacsong}), $\alpham = \alpha_{\text{T},\ox{}}$.

With knowledge of $\tp$ solved from Eq.~(\ref{eq:hp}), and provided an expression for $\mdotox$ derived from the particle kinetics (Section~\ref{sec:kinetics}), Eq.~(\ref{eq:bsheat}) and (\ref{eq:bsmass}) with associated relations can be solved numerically for $\{\tth, \, \coxp, \, \coxth \}$. The values of $\qdot$ and $\mdotox$ can then respectively be obtained from either side of Eq.~(\ref{eq:bsheat}) and  (\ref{eq:bsmass}), to obtain the heat and mass transport rates considering Knudsen effects.

\subsubsection{Explicit method}

The general boundary sphere implementation described by Eq.~(\ref{eq:bsheat}) and (\ref{eq:bsmass}) requires solving a coupled system of nonlinear equations with associated relations. Under the assumption of small temperature and concentration differences between the bulk gas and the particle surface, the heat transport rate can instead be expressed by applying a transitional correctional factor $\betat$ to the continuum rate. Liu et al. \cite{liu2006} derived such a formulation based on the Springer and Tsai model \cite{springer1965}: 
\beq \label{eq:betat}
\betat = \f{\qdot}{\qdotc} = \bigg(\f{1}{1 + \Kn} + \f{1}{2}\gt \Kn \bigg) ^ {-1}
\eeq
where $\qdot$ is the actual heat transport rate accounting for Knudsen transition transport effects, $\qdotc$ is the rate which would be obtained purely from continuum--Eq.~(\ref{eq:qdotc})--and $\gt = 8f /[\alphat(\gamma + 1)]$ is the geometry-dependent heat transfer factor. Analogously, a transition factor $\betam$ is derived for the mass transport rate $\mdotox$,
\beq \label{eq:betam}
\betam = \f{\mdotox}{\mdotc} = \bigg(\f{1}{1 + \Kn} + \f{1}{2} \gm \Kn \bigg) ^ {-1}
\eeq
where $\mdotc$ is obtained from Eq.~(\ref{eq:mdotc}), $\gm = 4f/(\alpham \gamma \Le)$ is the geometry-dependent mass transfer factor, and $\Le$ is the Lewis number. A detailed derivation of Eq.~(\ref{eq:betam}) is provided in Appendix \ref{sec:a1}. Equations~(\ref{eq:betat}) and (\ref{eq:betam}) are valid for arbitrary Kn and assume $\th = \mfpg$, where $\mfpg$ is obtained from Eq.~(\ref{eq:mfpg}).

%%%%%%%%%%%%%%%%%%%%%%%%%%%%%%%%%%%%%%%%%%%
% KINETICS
%%%%%%%%%%%%%%%%%%%%%%%%%%%%%%%%%%%%%%%%%%%

\subsection{Particle oxidation kinetics} \label{sec:kinetics}

\subsubsection{Parabolic model with compact oxide layers}

The parabolic kinetic model is based on the model of Mi et al \cite{mi2022}, wherein a multi-layered, compact oxide shell is formed on the surface of the iron core, as shown in Fig.~\ref{fig:oxidation}, through the parallel reactions:
\beqarr
\label{eq:parabolic1} \fe + \f{1}{2} \, \ox &\rightarrow& \feo \\
\label{eq:parabolic2} 3 \, \fe + 2 \, \ox &\rightarrow& \ftf. 
\eeqarr
The particle oxidation kinetics are formulated through a parabolic rate law, where the rate-limiting step is the internal diffusion rate of iron ions through the oxide layers. Consequently, the kinetics are independent of $\coxp$. The kinetic rate of formation of the oxide $i$ is,
\beq \label{eq:mdotik}
\mdotik = \rho_i A_i \f{\d X_i}{\dt} \equiv \rho_i A_i \bigg(\f{r_i - X_i}{r_i X_i}\bigg) k_{\infty,i} \exp \bigg( \f{-\tai}{ \tp} \bigg)
\eeq
where $\rho_i$ is the oxide solid-phase density, $A_i = 4 \pi r_i^2$ is the reaction surface area, $r_i$ is the reaction radius, $k_{\infty,i}$ is the pre-exponential factor, and $\tai$ is the activation temperature. The reaction kinetic parameters are provided in Table~\ref{tab:properties}. The reactions occur at the external surface of the oxide shells, such that the reaction radii are $\rfeo = \rfe + \xfeo$, and $\rftf = \rp$. The formulation for $\d X_i / \dt$ provided in Eq.~(\ref{eq:mdotik}) is adjusted from Ref.~\cite{mi2022} to take into account curvature effects in the transport rate of the ions. The kinetic rate of consumption of \fe{} and \ox{} can then be obtained through:
\beqarr
\label{eq:mdotfek} \mdotfek &=& \sum_i \nu_\f{\fe}{i} \mdotik \\
\mdotoxk &=& \sum_i \nu_\f{\ox}{i} \mdotik.
\eeqarr

Due to the independence of the kinetics on $\coxp$, the interplay between the kinetic- and diffusion- limited combustion regimes is captured through a switch-type model \cite{mi2022}. The maximum transport-limited consumption rate of \ox{}, $\mdotoxd$, is determined by setting $\coxp = 0$ in the Knudsen mass transport equations. The resulting rate is compared to $\mdotoxk$, and the lowest value is selected as the actual \ox{} consumption rate, $\mdotox = \min \{\mdotoxk, \, \mdotoxd\}$. If $\mdotoxk \leq \mdotoxd$, Eq.~(\ref{eq:mdotik}) and (\ref{eq:mdotfek}) can be used directly for the rate of change of the state variables. In the opposite case, the \ox{} is partitioned through reactions~(\ref{eq:parabolic1}) and (\ref{eq:parabolic2}) proportionally to the kinetic rates of each reaction, and the rates are adjusted as $\mdotid = (\mdotoxd / \mdotoxk) \mdotik$, and $\mdotfed = (\mdotoxd / \mdotoxk) \mdotfek$.

\subsubsection{First-order model with porous oxide layer}

The second kinetic model is based on the model of Hazenberg and van Oijen \cite{hazenberg2021}, which considers a single-step reaction:
\beq \label{eq:firstorder}
\fe + \f{1}{2} \, \ox \rightarrow \feo.
\eeq
The oxidizer consumption rate at the particle surface is formulated with a first-order Arrhenius rate law,
\beq \label{eq:mdotox1firstorder}
\mdotox = k_1 \coxp \afe \equiv \kinfone \exp \bigg(\f{-\taone}{\tp} \bigg) \coxp \afe
\eeq
where $\afe = 4 \pi \rfe^2$ is the reaction surface area, $k_1$ is the kinetic rate of the reaction, $\kinfone$ is the pre-exponential constant, and $\taone$ is the activation temperature. The reaction parameters are provided in Table~\ref{tab:properties}. The \feo{} oxide shell is assumed to be porous and to cause no hindrance on the transport of \ox{}, as shown in Fig.~\ref{fig:oxidation}, which results in the reaction surface area to be $\afe$. The Knudsen-corrected oxidizer transport rate is evaluated as,
\beq \label{eq:mdotox2firstorder}
\mdotox = \beta \betam (\coxg - \coxp) \afe 
\eeq
where $\beta = \D^* / \rfe$ is the diffusive velocity evaluated with the two-third law. The parameter $\betam$ can either be obtained directly from Eq.~(\ref{eq:betam}) in the explicit method, or it can be computed by solving the system defined by Eq.~(\ref{eq:bsheat}) and (\ref{eq:bsmass}), then computing $\betam = \mdotox / \mdotc$ in the implicit method. An effective diffusive velocity can then be defined as:
\beq
\betaeff = \betam \beta.
\eeq
The standard procedure for first-order reactions can then be applied, wherein Eq.~(\ref{eq:mdotox1firstorder}) and (\ref{eq:mdotox2firstorder}) are equated and solved for $\coxp$, and the result is substituted back in Eq.~(\ref{eq:mdotox1firstorder}) to obtain,
\beq \label{eq:mdotoxfinalfirstorder}
\mdotox = \bigg(\f{k_1}{k_1 + \betaeff}\bigg) \betaeff \coxg \afe \equiv \Dastar \betaeff \coxg \afe
\eeq
where the Knudsen-corrected normalized Damk\"{o}hler number $\Dastar = k_1 / (k_1 + \betaeff)$ has been defined. The consumption rate of Fe and the production rate of \feo{} can be related through stoichiometric coefficients using Eq.~(\ref{eq:firstorder}) and (\ref{eq:mdotoxfinalfirstorder}) to the consumption rate of \ox{}: $\mdotfe = \nu_\f{\fe}{\ox} \mdotox$, and $\mdotfeo = \nu_\f{\feo}{\ox} \mdotox$.

%%%%%%%%%%%%%%%%%%%%%%%%%%%%%%%%%%%%%%%%%%%
% GOVERNING EQUATIONS
%%%%%%%%%%%%%%%%%%%%%%%%%%%%%%%%%%%%%%%%%%%

\subsection{Governing equations and ignition criterion}  

The governing equations for the rate of change of the state variables are:
\beqarr
\label{eq:gov1} \f{\d \mfe}{\dt} &=& - \mdotfe \\
\label{eq:gov2} \f{\d m_i}{\dt} &=& \dot{m}_i \\
\label{eq:gov3} \f{\d \hp}{\dt} &=& \sum_i (\dot{m}_i q_i) + \mdotox h_\ox  - \qdot.
\eeqarr
On the right-hand-side of Eq.~(\ref{eq:gov3}), the first term represents the energy release in the particle from the formation of the oxides, where $q_i$ is the heating value of the oxides, provided in Table~\ref{tab:properties}; the second term relates to the enthalpy increase of the particle due to the incorporation of \ox{}, where $h_\ox{}$ is the enthalpy of \ox{} computed at $\tth$ with the implicit method, or at $\tg$ with the explicit method; and the third term is the Knudsen-corrected conductive heat loss rate from the particle surface to the surrounding gas mixture, obtained with the implicit or explicit method.

Given an initial particle and gas temperature, $T_\text{p,0} = \tg$; a bulk gas pressure, $p = 1~\text{atm}$; an initial particle diameter, $\dpzero$; and an initial oxide layer thickness, $X_0 = \sum_j X_{j,\text{0}}$, the governing equations are solved in time with the MATLAB solver \textit{ode15s}. In the case of the parabolic kinetic model, the initial oxide layer thickness $X_0$ is partitioned into 95\% FeO and 5\% \ftf{} by thickness \cite{mi2022}. In the case of the first-order model, $X_0 = 0$, as $X$ has no impact on the kinetic rate. Equations~(\ref{eq:gov1})--(\ref{eq:gov3}) are numerically integrated until particle burnout, or until the particle has undergone thermal runaway, which is the ignition criterion. For the first-order model, this translates to $\Dastar$ approaching unity.

\begin{table}
	\centering
	\caption{Properties and kinetic parameters for the iron particle ignition model.}
	\label{tab:properties}
	
	\begin{tabular}{|c|c|c|c|}
	
	\toprule[3pt]

	Description & Symbol & Value & Units \\
	
	\midrule[3pt]

	\multirow{3}{*}{Density}
	&$\rhofe$ & 7874 &   \\ 
	&$\rhofeo$ & 5745 & kg/m\super{3} \\ 
	&$\rhoftf$ & 5170 &  \\ \hline
	
	\multirow{2}{*}{Specific heating value}
	&$\qfeo$ & 3.787 &  \multirow{2}{*}{MJ/kg} \\ 
	&$\qftf$ & 4.841 &  \\ \hline
	
	\multirow{3}{*}{Kinetic constant}
	&$\kinfone$ & 7.50 x 10\super{6} \cite{hazenberg2021} & m/s \\ \cline{2-4} 
	&$k_{0,\feo}$ & 2.670 x 10\super{-4} \cite{mi2022} & \multirow{2}{*}{m\super{2}/s} \\ 
	&$k_{0,\ftf}$ & 1.027 x 10\super{-6} \cite{mi2022} & \\ \hline
	
	\multirow{3}{*}{Activation temperature}
	&$\taone$ & 14400 \cite{hazenberg2021} & K \\ \cline{2-4}
	&$T_{\text{a},\feo}$ & 20319 \cite{mi2022} & \multirow{2}{*}{K} \\
	&$T_{\text{a},\ftf}$ & 21310 \cite{mi2022} &  \\ 
	\hline
	
	\multirow{2}{*}{Diffusion volume} & $v_\ox$ & 16.6 \cite{fuller1966} & \multirow{2}{*}{-} \\
	& $v_\inert$  & 17.9 \cite{fuller1966} & \\ 
	
	\bottomrule[3pt]
	
	\end{tabular}
	
\end{table}

%%%%%%%%%%%%%%%%%%%%%%%%%%%%%%%%%%%%%%%%%%%
% STEADY-STATE MODEL
%%%%%%%%%%%%%%%%%%%%%%%%%%%%%%%%%%%%%%%%%%%

\subsection{Steady-state ignition model formulation} \label{sec:steadyformulation}

The unsteady model is compared to a simple steady-state Semenov analysis, not considering the growth of the oxides, with the Knudsen correction factors obtained from the explicit boundary sphere method. Using Eq.~(\ref{eq:qdotc}) and (\ref{eq:betat}), the heat loss rate from the particle can be expanded to:
\beq \label{eq:heatloss}
\qdot = 8 \pi \rp^2 k^* (\tp - \tg) \bigg(\f{\rp + \th}{2 \rp^2 + \gt \th \rp + \gt \th^2} \bigg)
\eeq
where $\th \equiv \mfpg$. The heat generation rate in the particle from the formation of the oxides is $\qdotr = \sum_i \dot{m}_i q_i $. In the parabolic kinetic model, using Eq.~(\ref{eq:mdotik}) results in,
\beq \label{eq:qdotrparabolic}
\qdotr = \tilde{q}_\feo \f{(\rp - X_0 \delta_\ftf)(\rp - X_0)}{(1-\delta_\ftf) X_0} + \tilde{q}_\ftf \f{(\rp - X_0 \delta_\ftf)\rp }{X_0 \delta_\ftf}
\eeq
where $\tilde{q}_j = 4 \pi q_j \rho_j k_{0,j} \exp(-T_{\text{a},j}/\tp)$, and $\delta_\ftf = 0.05$ is the initial proportion of \ftf{} in the oxide layer. In the case of first-order kinetics, Eq.~(\ref{eq:betam}), (\ref{eq:firstorder}), and (\ref{eq:mdotoxfinalfirstorder}) can be used to show:
\beq \label{eq:qdotrfirstorder}
\qdotr = 8 \pi \rp^2 \D^* \coxg \Big( \nu_\f{\feo}{\ox} \qfeo k_1 \Big) \bigg( \f{\rp + \th}{ \big[ 2 \rp^2 + \gm \rp \th + \gm \th^2 \big] k_1 + 2 \D^* (\rp + \th)} \bigg).
\eeq
Semenov ignition occurs when $\dot{Q} = \qdotr$ and $\d\qdot/\d \tp = \d \qdotr/\d \tp$. The criteria are solved as a function of $\dpzero$ and compared to the unsteady results.

%\bibliographystyle{../pci}
%\bibliography{../afl_refs}

\section{Results and analysis} \label{sec:results}

The parabolic and first-order kinetic models are two independent models for the ignition of an iron particle. As previously stated, the parabolic kinetics are proposed to provide a more realistic representation of the high-temperature solid-phase oxidation of iron, and the results obtained with this model are presented independently in Section~\ref{sec:results_parabolic}. The first-order model results are then presented as a comparison to the parabolic model in Section~\ref{sec:results_firstorder}. Unless otherwise stated, the explicit method is used to compute the boundary sphere transport rates (Section~\ref{sec:balanceknudsen}), since a small temperature and oxidizer concentration difference between the bulk gas and the particle surface are assumed in the pre-ignition phase. In some instances where this assumption is not valid (Section~\ref{sec:burningregime}), the implicit method is used. 
%Appendix~\ref{sec:a2} provides an analysis which validates the explicit method is accurate to resolve the ignition problem, when compared to the implicit method. 

\subsection{Parabolic kinetics} \label{sec:results_parabolic}

%%%%%%%%%%%%%%%%%%%%%%%%%%%%%%%%%%%%%%%%%%%
% SAMPLE
%%%%%%%%%%%%%%%%%%%%%%%%%%%%%%%%%%%%%%%%%%%

\subsubsection{Sample results - transient behavior and ignition} 

\begin{figure}[h]
	\centering 
	
	\begin{subfigure}{.32\linewidth}
	\centering 
	\includegraphics[width=\linewidth]{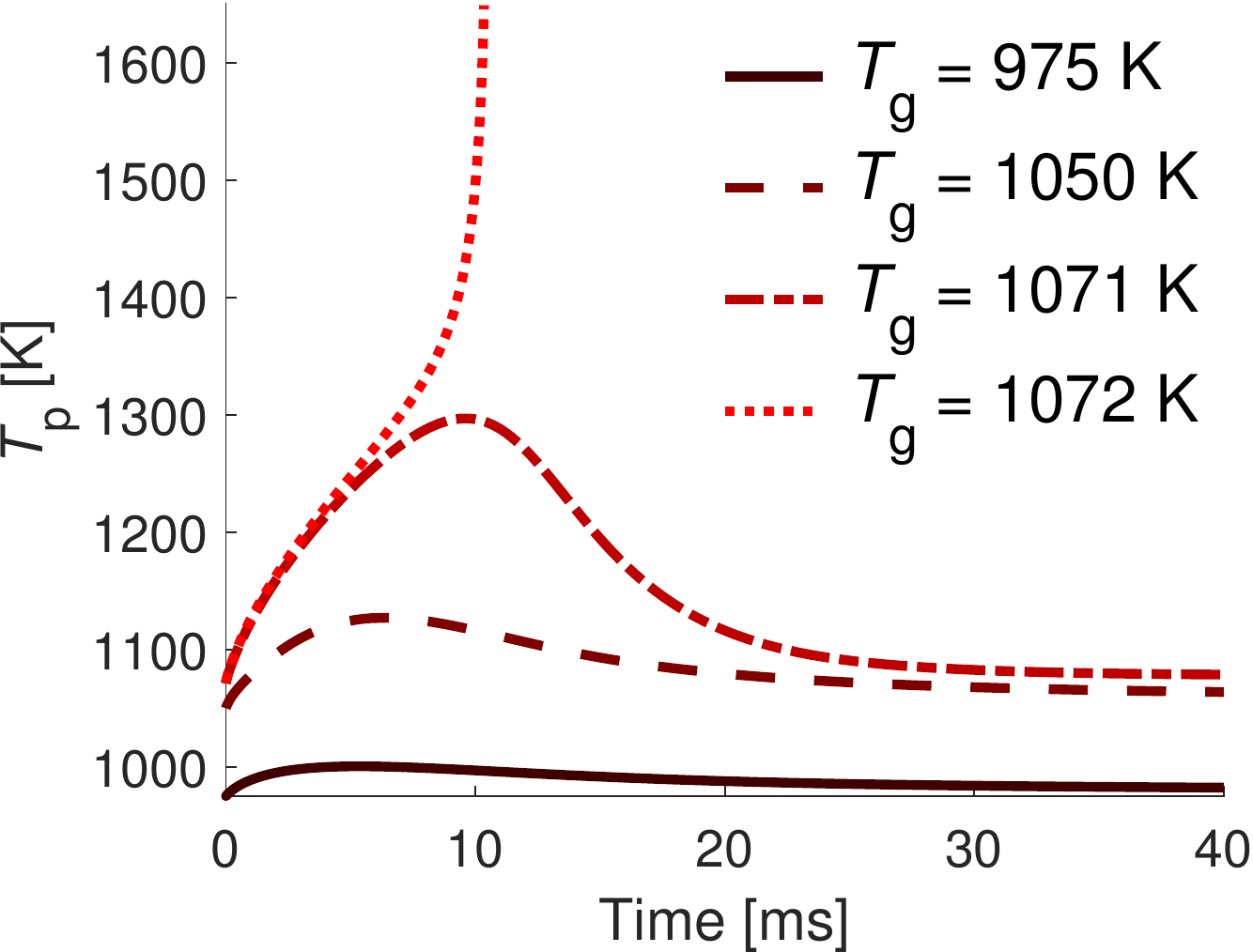}
	\caption{Particle temperature.}
	\label{fig:sampleT_parabolic_10nm} 
	\end{subfigure}
	\begin{subfigure}{.32\linewidth}
	\centering 
	\includegraphics[width=\linewidth]{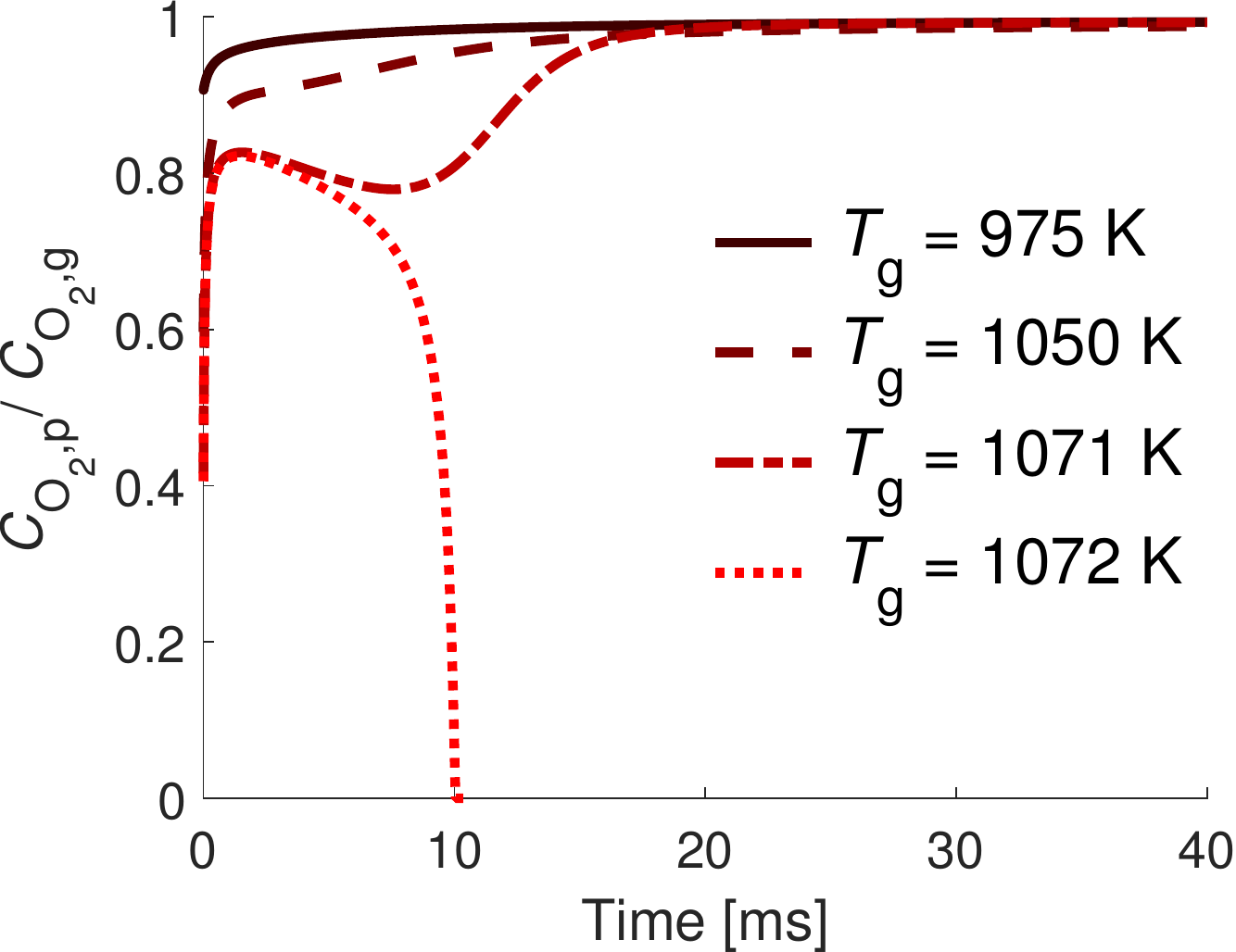}
	\caption{Normalized oxidizer concentration.}
	\label{fig:sampleC_parabolic} 
	\end{subfigure}
	\begin{subfigure}{.32\linewidth}
	\centering 
	\includegraphics[width=\linewidth]{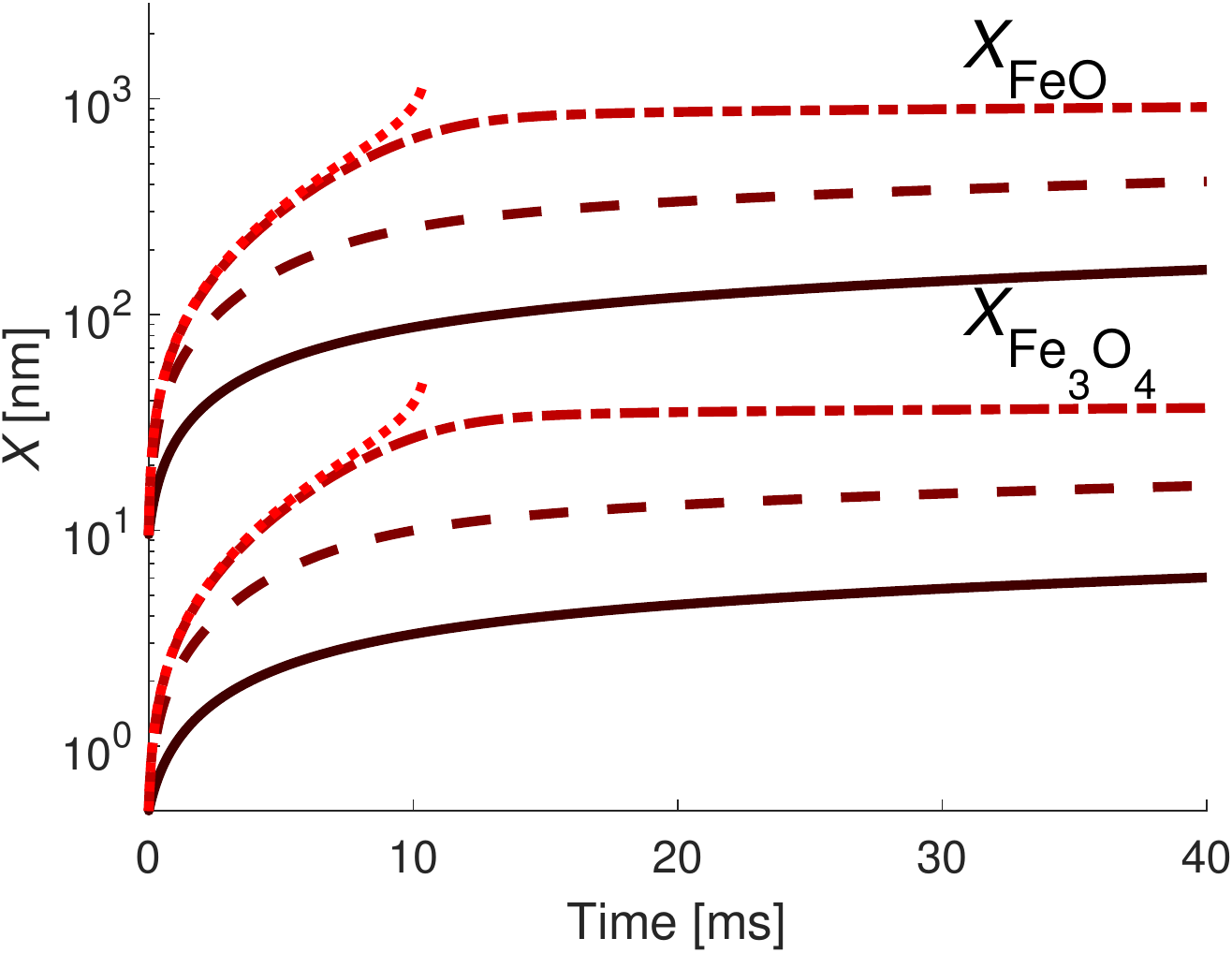}
	\caption{Growth of the oxide layers.}
	\label{fig:sampleX_parabolic} 
	\end{subfigure}
	
	\caption{Particle transient behavior for $\dpzero = 20~\upmu$m, $X_0 =$ 10 nm at different $\tg$ (parabolic kinetics). Ignition occurs at $\tg = 1072$~K.}
	\label{fig:sample_parabolic} 
	
\end{figure}

Figure~\ref{fig:sample_parabolic} shows the particle temperature, $\tp$, normalized oxidizer concentration at the particle surface, $\coxp/\coxg$, and growth of the oxide layers, $X_i$, for a burning particle placed in different bulk gas temperatures, $\tg$, with the parabolic kinetic model. As $\tg$ is increased, $\tp$ increasingly separates from the bulk gas as seen in Fig.~\ref{fig:sampleT_parabolic_10nm}, which accelerates the particle kinetics. However, the growth of the oxides shown in Fig.~\ref{fig:sampleX_parabolic} has the opposite effect, while also leading to additional inert thermal mass to be heated by the energy release of the particle. As well, the heat loss rate from the particle increases with the temperature separation $\Delta T = \tp - \tg$. For $\tg$ below the critical ignition temperature $\tign$, $\tp$ returns close to $\tg$ after reaching a peak; the particle burns out in the kinetic-limited regime, and $\coxp/\coxg$ tends to one. At the critical gas temperature for ignition, the exponential dependence of the kinetics on $\tp$ exceeds the adverse effect of the oxide layer growth and heat loss rate. The particle undergoes thermal runaway and transitions to the diffusion-limited regime, and $\coxp$ reaches zero, as seen in Fig.~\ref{fig:sampleC_parabolic} for $\tg = 1072$~K. 

%%%%%%%%%%%%%%%%%%%%%%%%%%%%%%%%%%%%%%%%%%%
% DP0 AND X0
%%%%%%%%%%%%%%%%%%%%%%%%%%%%%%%%%%%%%%%%%%%

\subsubsection{Ignition behavior - effect of particle size and initial oxide layer thickness}

\begin{figure}
	\centering
	
	\begin{subfigure}{0.35\linewidth}
	\centering
	\includegraphics[width=\linewidth]{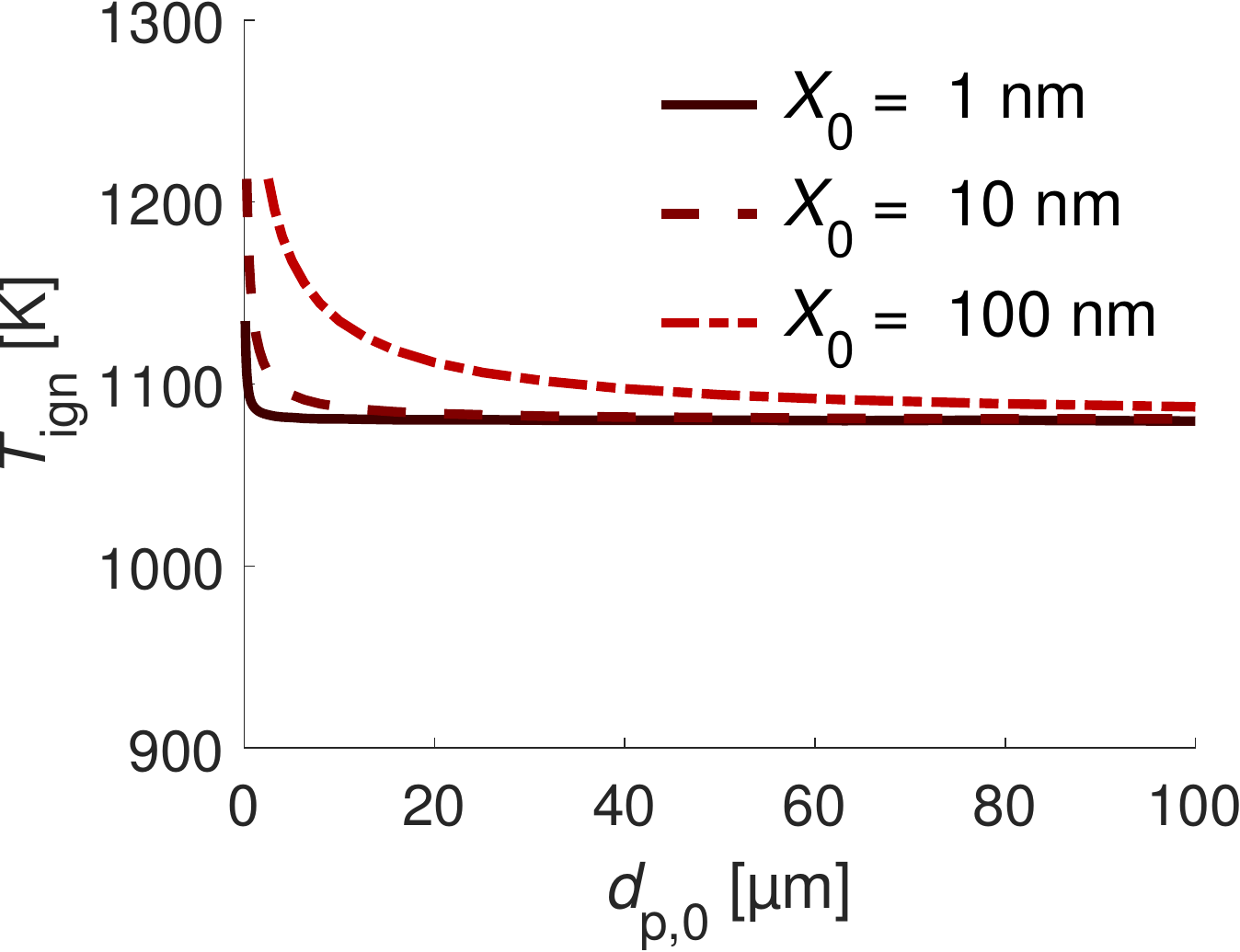}
	\caption{Continuum approximation.}
	\label{fig:tignvsdp0_parabolic_co}
	\end{subfigure}
	\begin{subfigure}{0.35\linewidth}
	\centering
	\includegraphics[width=\linewidth]{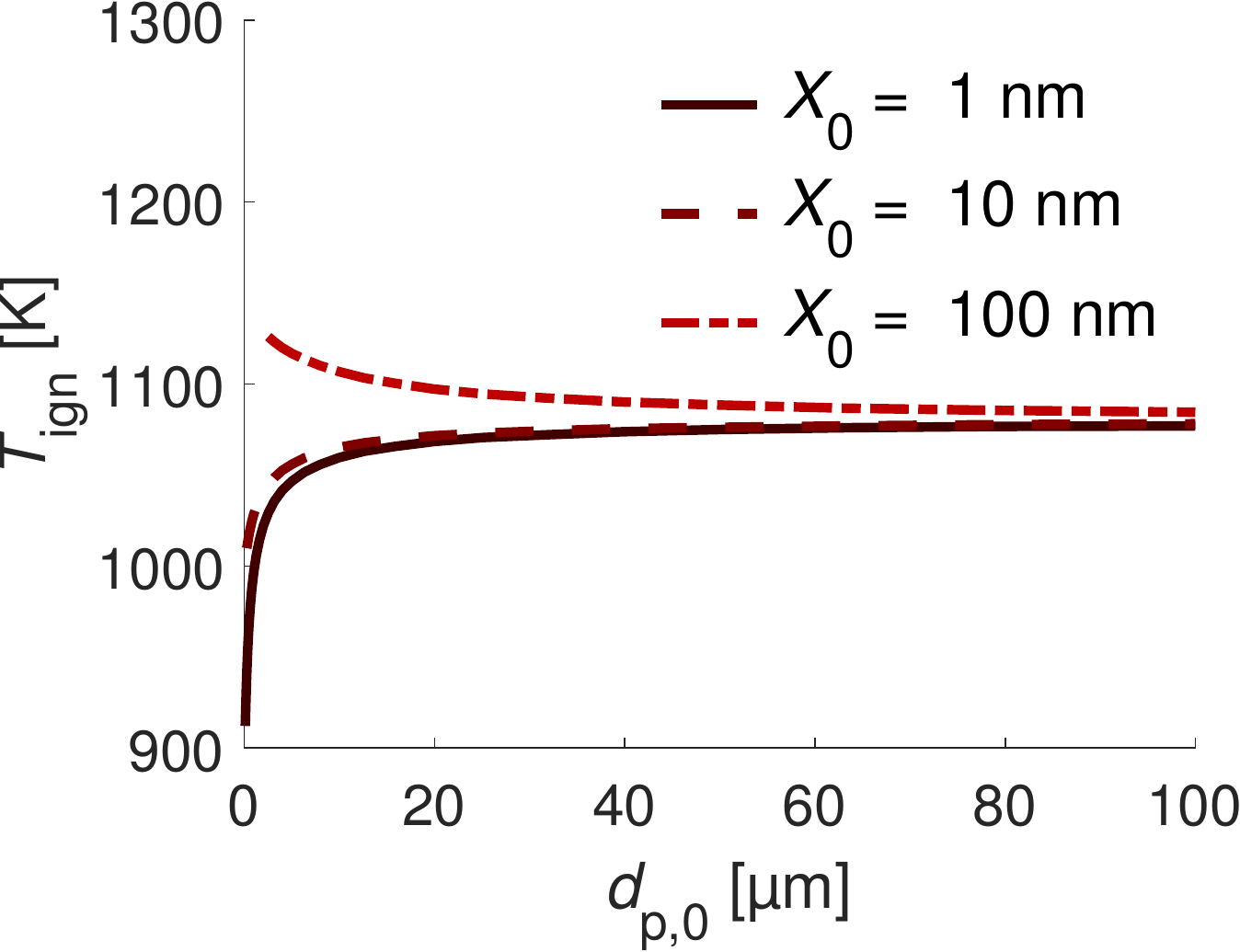}
	\caption{Knudsen modeling.}
	\label{fig:tignvsdp0_parabolic_knu}
	\end{subfigure}
	
	\caption{Ignition temperature as a function of initial particle size for different initial oxide layer thicknesses (parabolic kinetics).}
	\label{fig:tignvsdp0_parabolic}
	
\end{figure}

The critical gas ignition temperature, $\tign$, is solved as a function of particle size, $\dpzero$, and initial oxide layer thickness, $X_0$, as shown in Fig.~\ref{fig:tignvsdp0_parabolic}. The continuum and transition analyses are both conducted; the former results agree with those presented in Ref.~\cite{mi2022}. A physical interpretation of the plateau in the large-particle limit is provided in Ref.~\cite{mi2022}: as $\dpzero$ increases, the heat generation rate increases faster with particle size than the heat removal rate, facilitating particle ignition. However, larger particles grow a thicker oxide layer, which impedes ignition. The competition between these two effects eventually leads to an independence of $\tign$ on $\dpzero$. At large $\dpzero$, the transition analysis converges to the continuum results, as expected. In the small-particle limit, different trends are observed. For $X_0 =$ 1~and~10~nm, $\tign$ decreases with $\dpzero$, opposite to what is predicted by the continuum model. For $X_0 =$ 100~nm, $\tign$ increases with $\dpzero$, but the asymptotic behavior is weaker than its continuum counterpart.

The ability of the particle to ignite is dependent upon its ability to retain heat; hence, an important analysis of heat transfer processes as a function of particle size arises. In the free-molecular limit, where the limiting step for heat transfer is the low molecule-particle collision rate, the heat loss rate scales with the surface area available for collisions. In fact,
\beqarr
\rp + \theta &\rightarrow& \theta \\ 
\rp^2 + \gt \theta \rp + \gt \theta^2 &\rightarrow& \gt \theta^2 \\
\label{eq:qdot_small_limit} \text{Equation}~(\ref{eq:heatloss}) \Rightarrow \qdot &\rightarrow& \f{8 \pi k^*}{\gt \theta} (\tp - \tg) \rp^2
\eeqarr
which can be shown to be equivalent to Eq.~(\ref{eq:qdotfm}) and scales with $\rp^2$. Conversely, in the continuum limit, where the inter-molecular collisions hinder heat transfer, the heat loss rate scales with the distance over which this resistance is significant, which is the thermal boundary layer, that has a thickness proportional to $\rp$. In fact, 
\beqarr
\rp + \theta &\rightarrow& \rp \\
2 \rp^2 + \gt \theta \rp + \gt \theta^2 &\rightarrow& 2 \rp^2 \\
\label{eq:qdot_large_limit} \text{Equation}~(\ref{eq:heatloss}) \Rightarrow \qdot &\rightarrow& 4 \pi k^* (\tp-\tg) \rp  
\eeqarr
which is identical to Eq.~(\ref{eq:qdotc}) and scales with $\rp$. The Knudsen number across the transition regime quantifies the interplay between the free-molecular and continuum heat transfer limiting processes.

The trends observed in Fig.~\ref{fig:tignvsdp0_parabolic_knu} for small particles are then explained as follows. As $\dpzero$ decreases, the molecule-particle collision rate decreases. Since parabolic iron kinetics are independent of the delivery rate of oxidizer to the particle surface ($\coxp$), the internal reaction rates are not affected. However, the decreasing collision rate results in a decrease of the heat loss rate from the particle. This thermal insulating effect increases the ignition propensity of smaller particles as observed with $X_0 =$ 1~nm and $X_0 =$ 10~nm, reducing $\tign$. However, as $\dpzero$ decreases at constant $X_0$, the iron content available to react and generate heat is reduced, while the proportion of inert oxide thermal mass, which must be heated to undergo thermal runaway, increases. For a sufficiently large $X_0$ (e.g. 100~nm), the latter effect outweighs the thermal insulating effect of the transition transport regime, and the net result is an increase of $\tign$ with decreasing $\dpzero$.

It is of interest to note that, in the boundary sphere approach, a limiting sphere beyond which continuum appropriately describes transport processes is defined. The thermal boundary layer resistance around the limiting sphere scales with its radius $\rp + \theta$. In the free-molecular and continuum limits, only one of these terms dominate. Hence, in the free-molecular limit, the thermal boundary layer thickness scales with the gas molecular mean free path, while in the continuum limit, it scales with particle size.

%%%%%%%%%%%%%%%%%%%%%%%%%%%%%%%%%%%%%%%%%%%
% STEADY / UNSTEADY
%%%%%%%%%%%%%%%%%%%%%%%%%%%%%%%%%%%%%%%%%%%

\subsubsection{Steady and unsteady analyses - effect of oxide layer growth} \label{sec:steady_parabolic}

\begin{figure}
    \centering
    \includegraphics[width = 0.4\linewidth]{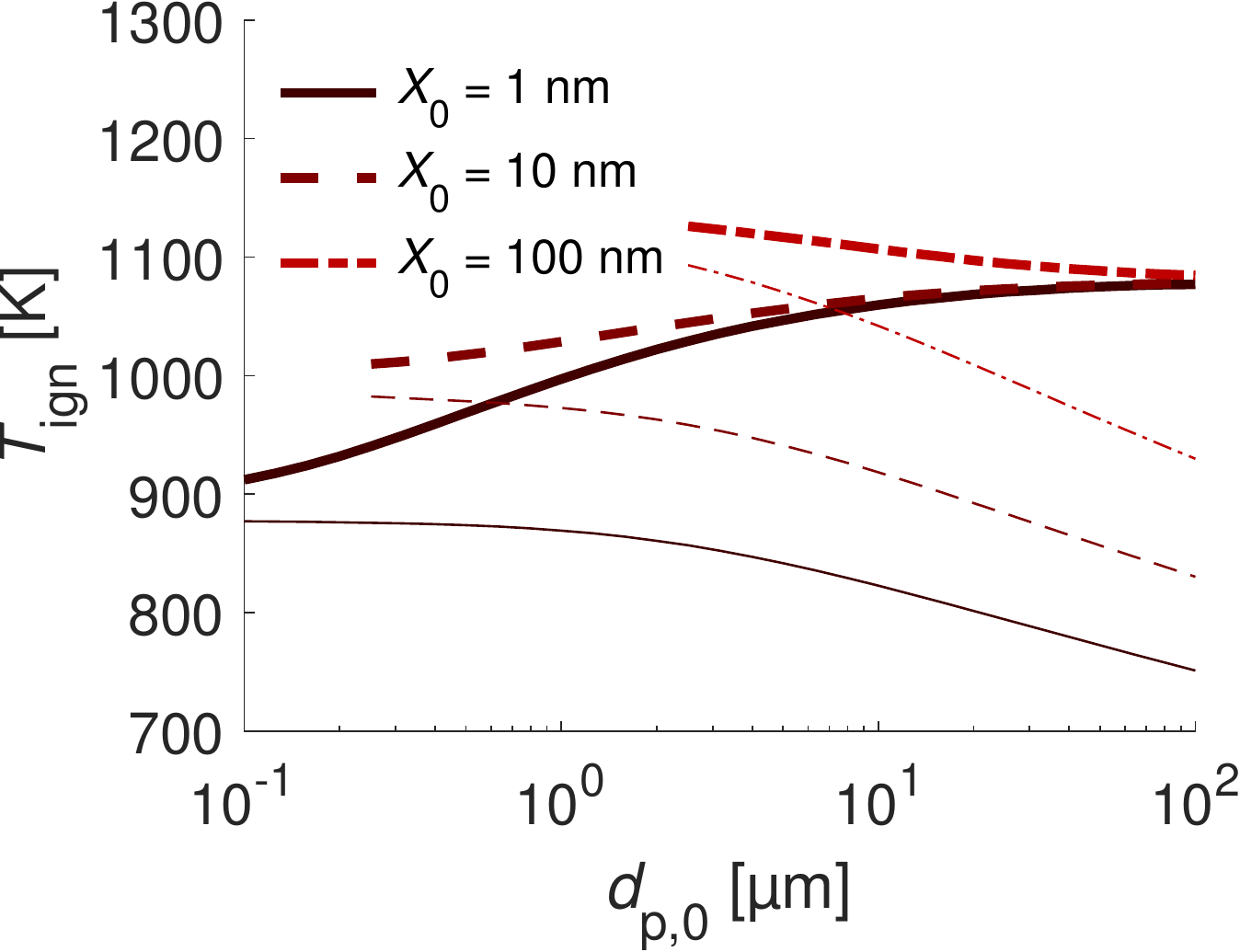}
    \caption{Comparison of the prediction of $\tign$ vs $\dpzero$ as resolved by the unsteady model--thick upper branches--and the steady model--thin lower branches (parabolic kinetics).}
    \label{fig:steady_parabolic}
\end{figure}

\begin{figure}
	\centering
	
	\begin{subfigure}{0.35\linewidth}
	\centering
	\includegraphics[width=\linewidth]{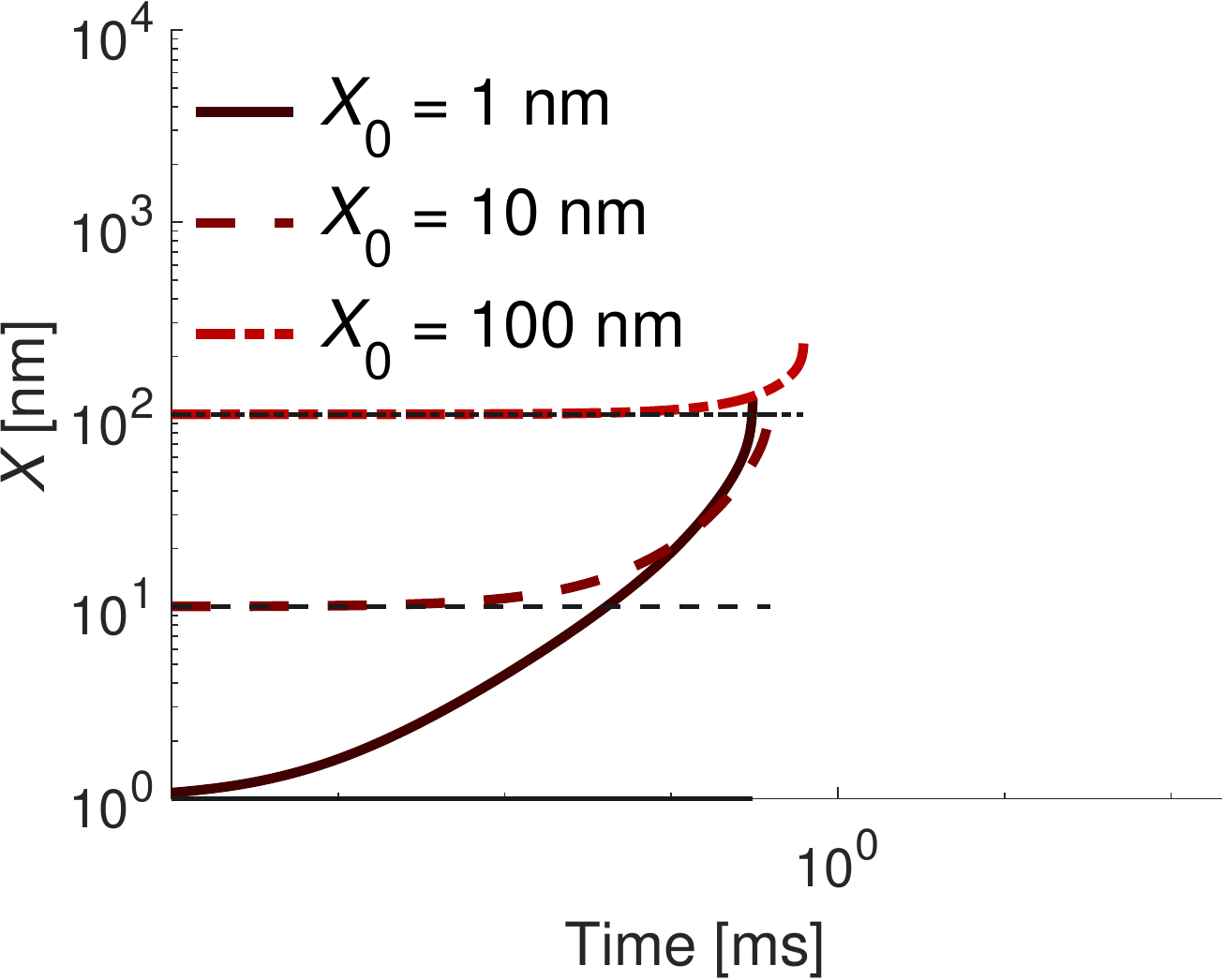}
	\caption{Small particle: $\dpzero = 2.51~\upmu$m.}
	\label{fig:oxidesmall} 
	\end{subfigure}
	\begin{subfigure}{0.35\linewidth}
	\centering
	\includegraphics[width=\linewidth]{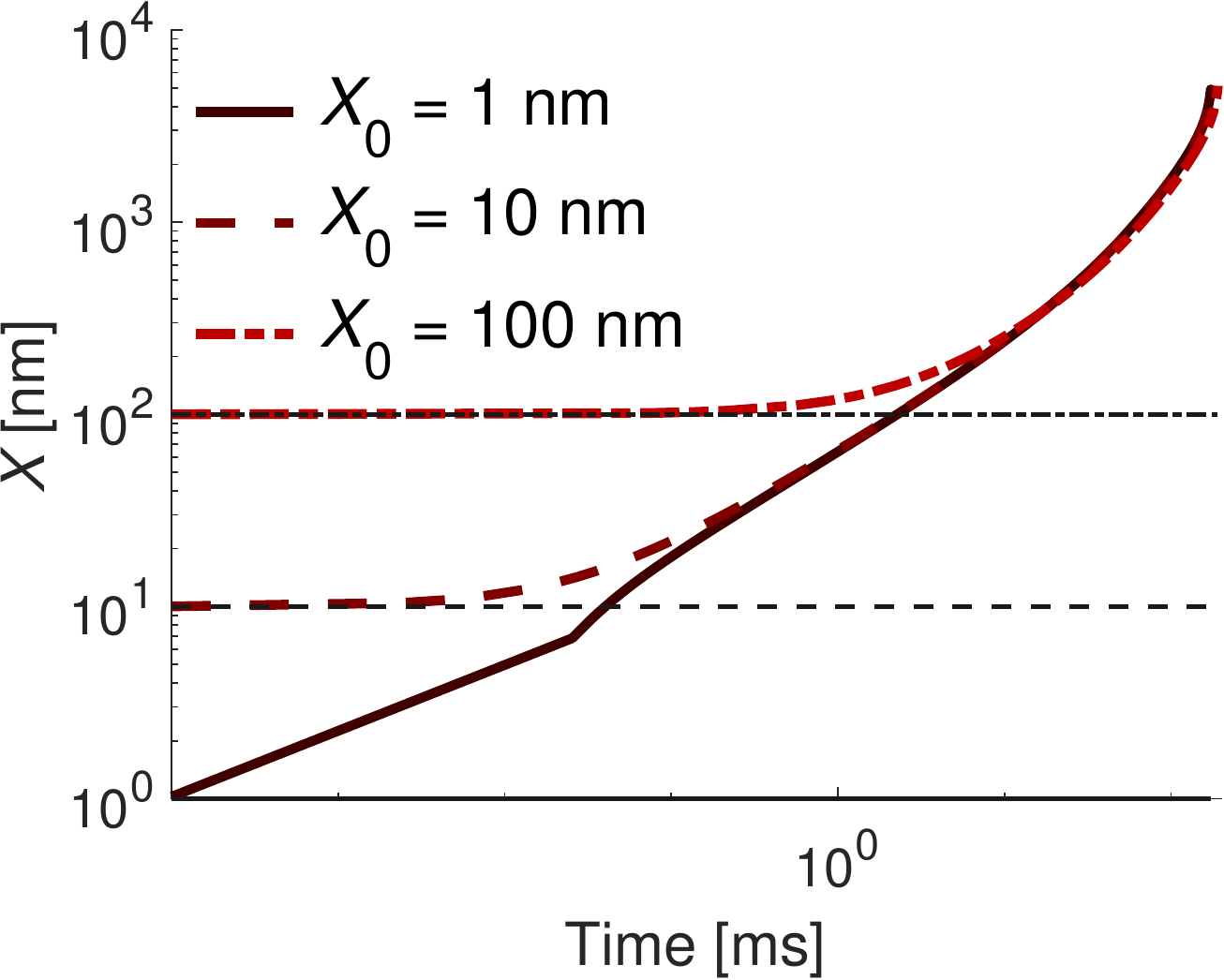}
	\caption{Large particle: $\dpzero = 100~\upmu$m.}
	\label{fig:oxidelarge} 
	\end{subfigure}
	
	\caption{Growth of the oxide scale over time at the critical ignition temperature (parabolic kinetics).}
	\label{fig:oxide}
	 
\end{figure}

The unsteady model is compared to a simple steady-state Semenov analysis as formulated in Section~\ref{sec:steadyformulation}. Figure~\ref{fig:steady_parabolic} shows the steady analysis always under-predicts $\tign$, since it neglects the oxide growth, which decreases the particle reaction rates. The discrepancy becomes larger with increasing $\dpzero$, while it decreases with increasing $X_0$. 
To understand this behavior, Fig.~\ref{fig:oxide} shows the growth of the oxide scale in time at $\tign$ for a small (\ref{fig:oxidesmall}, $\dpzero = 2.51~\upmu$m) and a large (\ref{fig:oxidelarge}, $\dpzero = 100~\upmu$m) particle, given different values of $X_0$. In Fig.~\ref{fig:oxidesmall}, the growth ranges from approximately one half to two orders of magnitude, while in Fig.~\ref{fig:oxidelarge}, it ranges from two to four orders of magnitude. Hence, the growth of the oxide scale before ignition ($\Delta X$) increases with $\dpzero$, which implies a larger discrepancy between the unsteady and steady models. As well, Fig.~\ref{fig:oxide} shows $\Delta X$ is larger for smaller values of $X_0$, which once again leads to larger discrepancies between the unsteady and steady ignition models. Figure~\ref{fig:oxidelarge} shows $X$ converges to approximately the same value at the moment of ignition for the large particle, independent of $X_0$, while the values differ for the small particle. Hence, $\tign$ becomes independent of $X_0$ in the large-particle limit, as shown by the converging curves in Fig.~\ref{fig:tignvsdp0_parabolic}, since the total mass content that must be heated to undergo thermal runaway remains approximately constant. In general, the large discrepancies between the unsteady and steady models, namely for large particles and small initial oxide layer thicknesses, imply that an unsteady analysis should be used to predict $\tign$ for kinetics featuring oxide layer growth and an adverse effect of the oxide layer on the kinetic reaction rates.

%%%%%%%%%%%%%%%%%%%%%%%%%%%%%%%%%%%%%%%%%%%
% CONTINUUM LIMIT
%%%%%%%%%%%%%%%%%%%%%%%%%%%%%%%%%%%%%%%%%%%

\subsubsection{Continuum approximation limit}

\begin{figure}

	\centering
	
	\begin{subfigure}{0.45\linewidth}
	\centering
	\includegraphics[width=\linewidth]{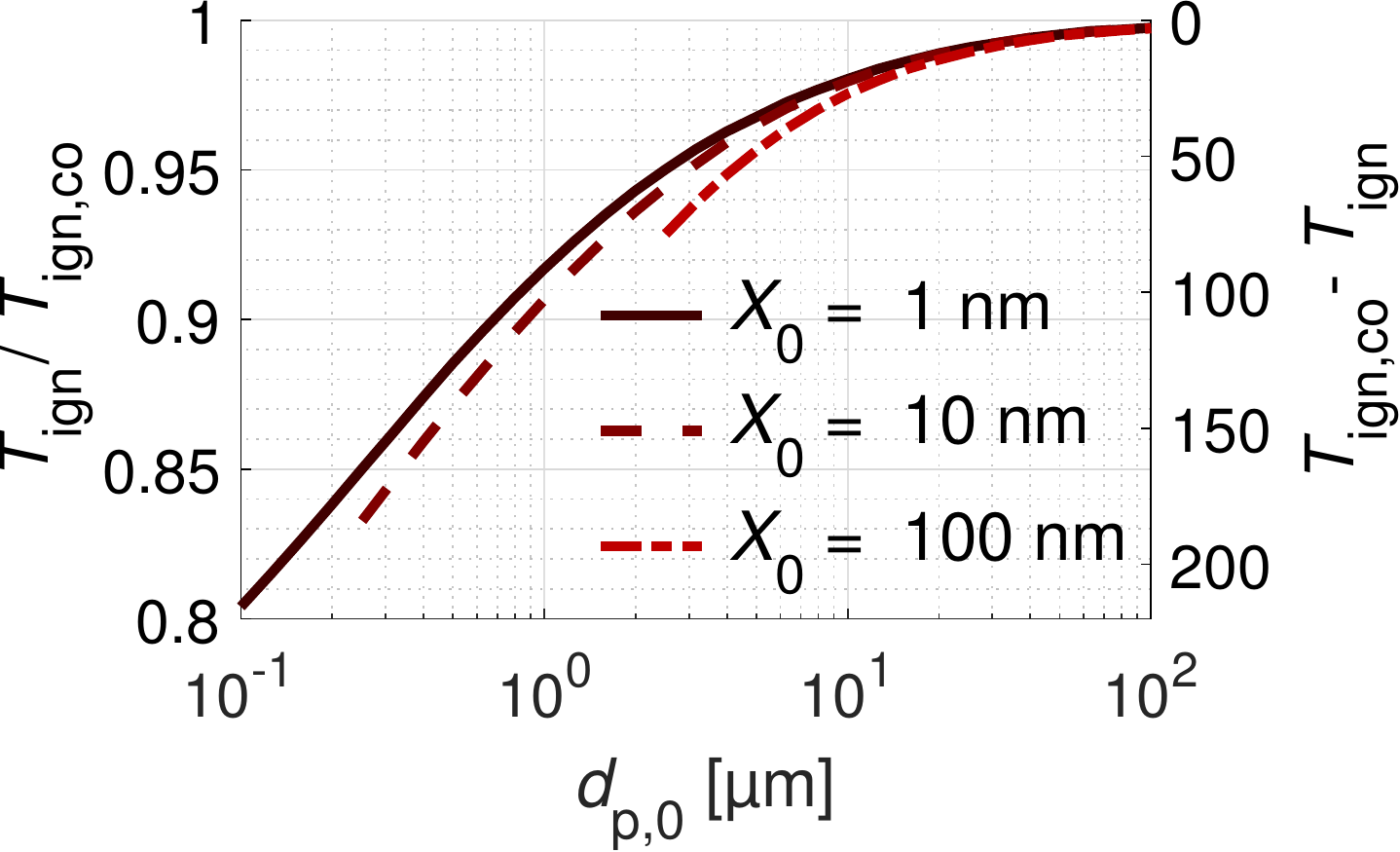}
	\caption{Effect of $X_0$ with correlation for $\alphat$.}
	\label{fig:compx0}
	\end{subfigure} \hfill
	\begin{subfigure}{0.45\linewidth}
	\centering
	\includegraphics[width=\linewidth]{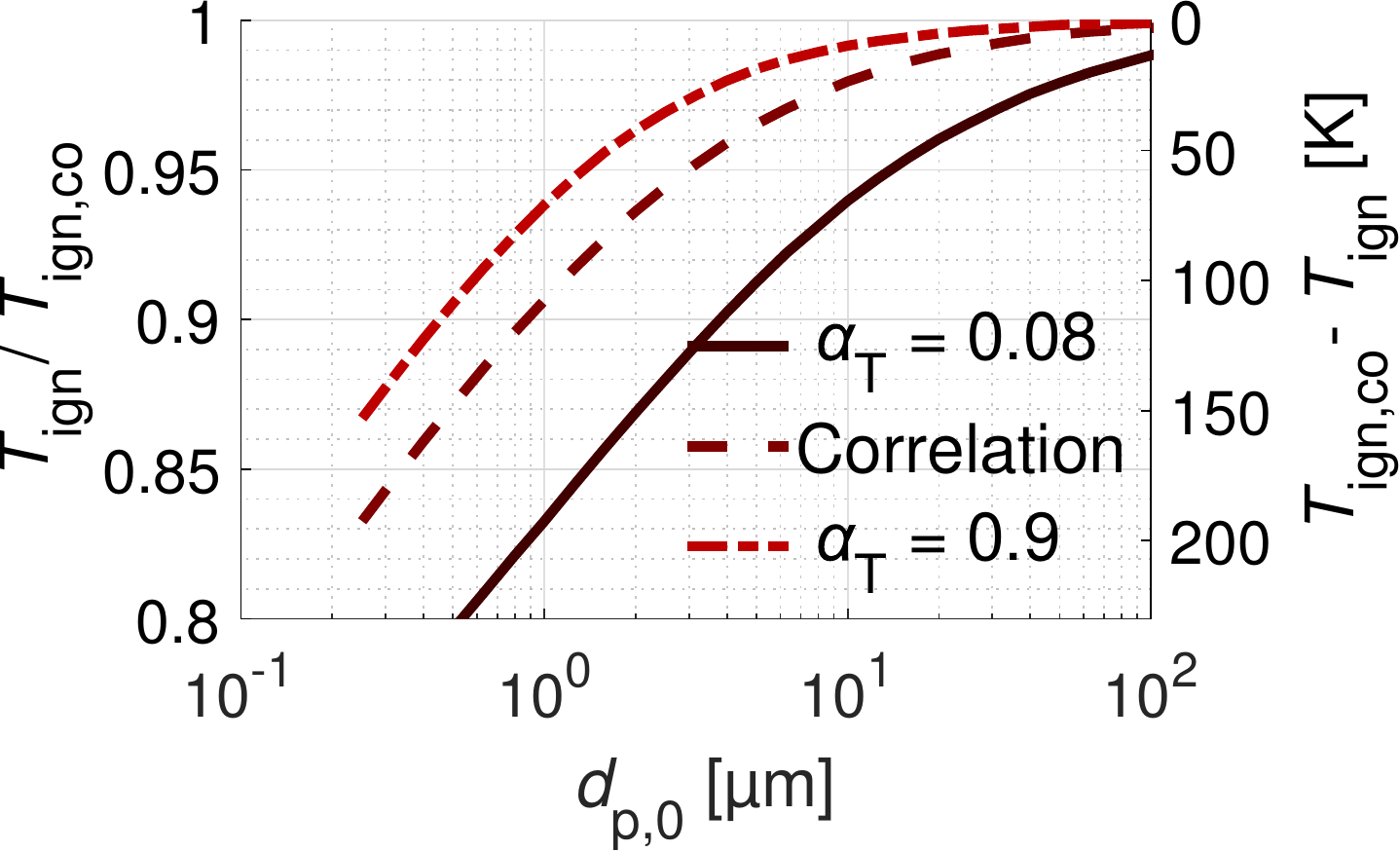}
	\caption{Effect of $\alphat$ in the confidence interval with $X_0 =$ 10 nm.}
	\label{fig:comptac}
	\end{subfigure}
	
	\caption{Comparison of the ignition temperature as a function of particle size with the transition ($\tign$) and continuum ($T_\text{ign,co}$) transport models (parabolic kinetics).}
	
	\label{fig:contlimit}
\end{figure}

Figure~\ref{fig:contlimit} compares $\tign$ to $T_\text{ign,co}$ as a function of $\dpzero$, where $T_\text{ign,co}$ are the continuum-predicted ignition temperatures. As expected, $T_\text{ign}/T_\text{ign,co} \rightarrow 1$ as $\dpzero$ increases. With the Song and Yovanovich \cite{song1987} correlation for $\alphat$, Fig~\ref{fig:compx0} shows continuum modeling predicts the same results as transition modeling to within 95\% for $\dpzero \gtrsim~4~\upmu$m, and to within 99\% for $\dpzero \gtrsim 30~\upmu$m, which corresponds to temperature differences of 10--50~K. In practice, such small differences cannot be resolved by current experimental measurement methods. 

The parameter $\alphat$ is a key variable governing $\tign$ in the transition and free-molecular regimes. There is large uncertainty in reported values of $\alphat$ for iron surfaces. Mohan et al. \cite{mohan2008} report $\alphat$ should be within the range 0.50 to 0.90 for metal surfaces. Literature for different iron-gas pairs reports $\alphat$ scattered between 0.03 and 0.64 \cite{daun2013, sipkens2014, sipkens2015, sipkens2015b}. In the temperature range 900--1650~K, the Song and Yovanovich \cite{song1987} correlation results in values of $\alphat$ ranging between 0.34--0.46 for a \ftt-air system, and between 0.56--0.63 for a Fe-air system. This significant uncertainty renders the appropriate selection of $\alphat$ difficult, and the impact of this uncertainty is shown in Fig.~\ref{fig:comptac}. The Song and Yovanovich \cite{song1987} correlation results are compared to the range 0.08--0.90 representative of the scattered data in the literature for Fe-N\sub{2} systems. Smaller $\alphat$ results in lower $\tign$, since the heat is less effectively carried away from the particle surface. At $\dpzero = 30~\upmu$m, the difference between continuum and transition transport modeling increases from 1\% to 3\%, or from 10 to 30~K, which still represents experimentally unresolvable variations of the ignition temperature. Reasonable values of $\alphat$ therefore yield no impact on the overall conclusion of the onset of transition effects being significant only in the low tens of microns. In general, the continuum modeling accuracy limit for iron particle ignition problems can be established at 30 $\upmu$m, which agrees with the order of magnitude reported by the majority of previous researchers in heterogeneous reaction problems \cite{ermoline2018, gopalakrishnan2011, mohan2008, shpara2020}.

%%%%%%%%%%%%%%%%%%%%%%%%%%%%%%%%%%%%%%%%%%%
% PEAK TEMPERAUTRE
%%%%%%%%%%%%%%%%%%%%%%%%%%%%%%%%%%%%%%%%%%%

\subsubsection{Post-ignition burning regime - effect of oxidizer concentration} \label{sec:burningregime}

\begin{figure}
	\centering
	
	\begin{subfigure}{0.45\linewidth}
	\centering
	\includegraphics[width=\linewidth]{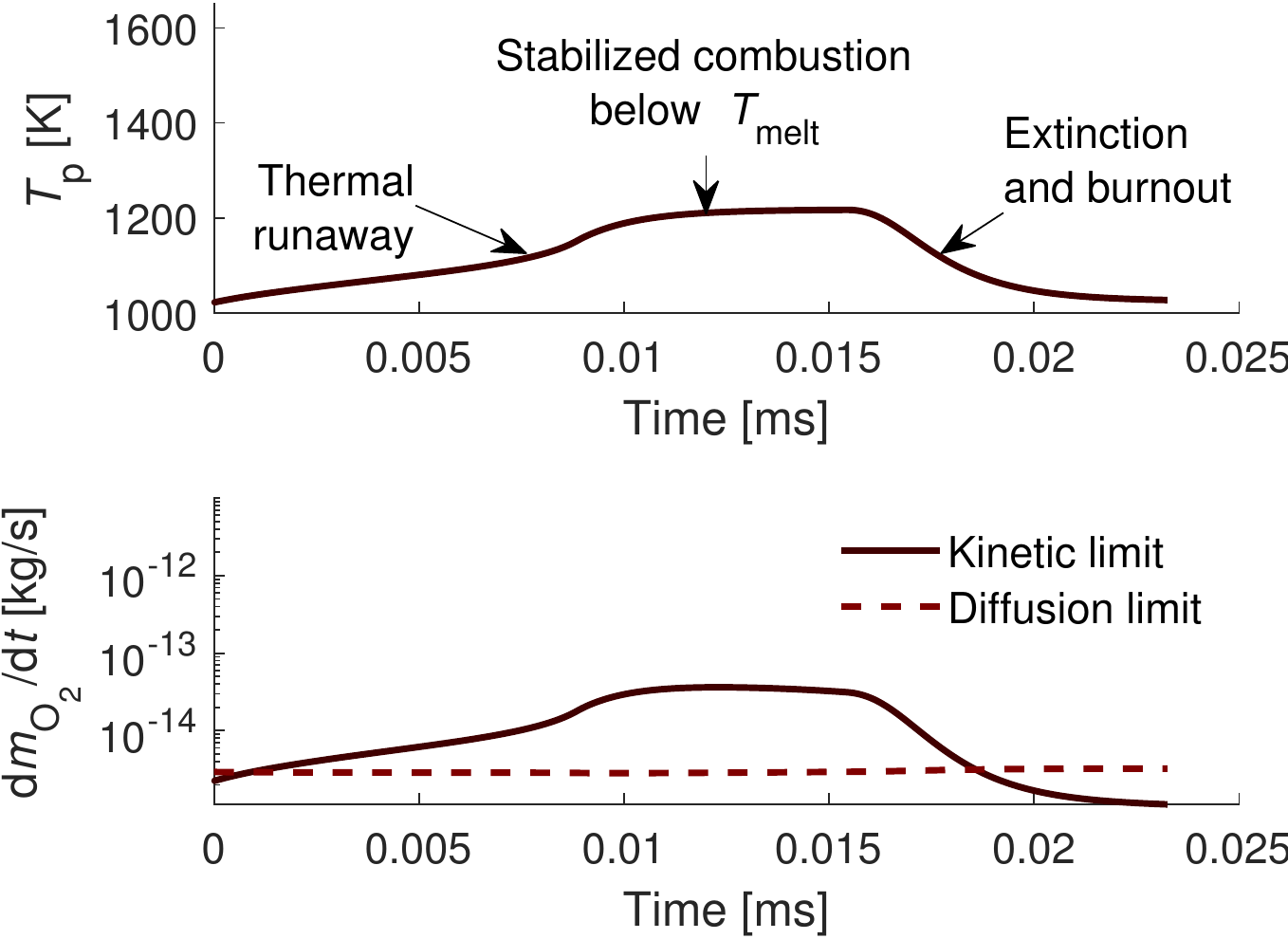}
	\caption{1\% oxidizer concentration.}
	\label{fig:sample_regime_1}
	\end{subfigure}
	\begin{subfigure}{0.395\linewidth}
	\centering
	\includegraphics[width=\linewidth]{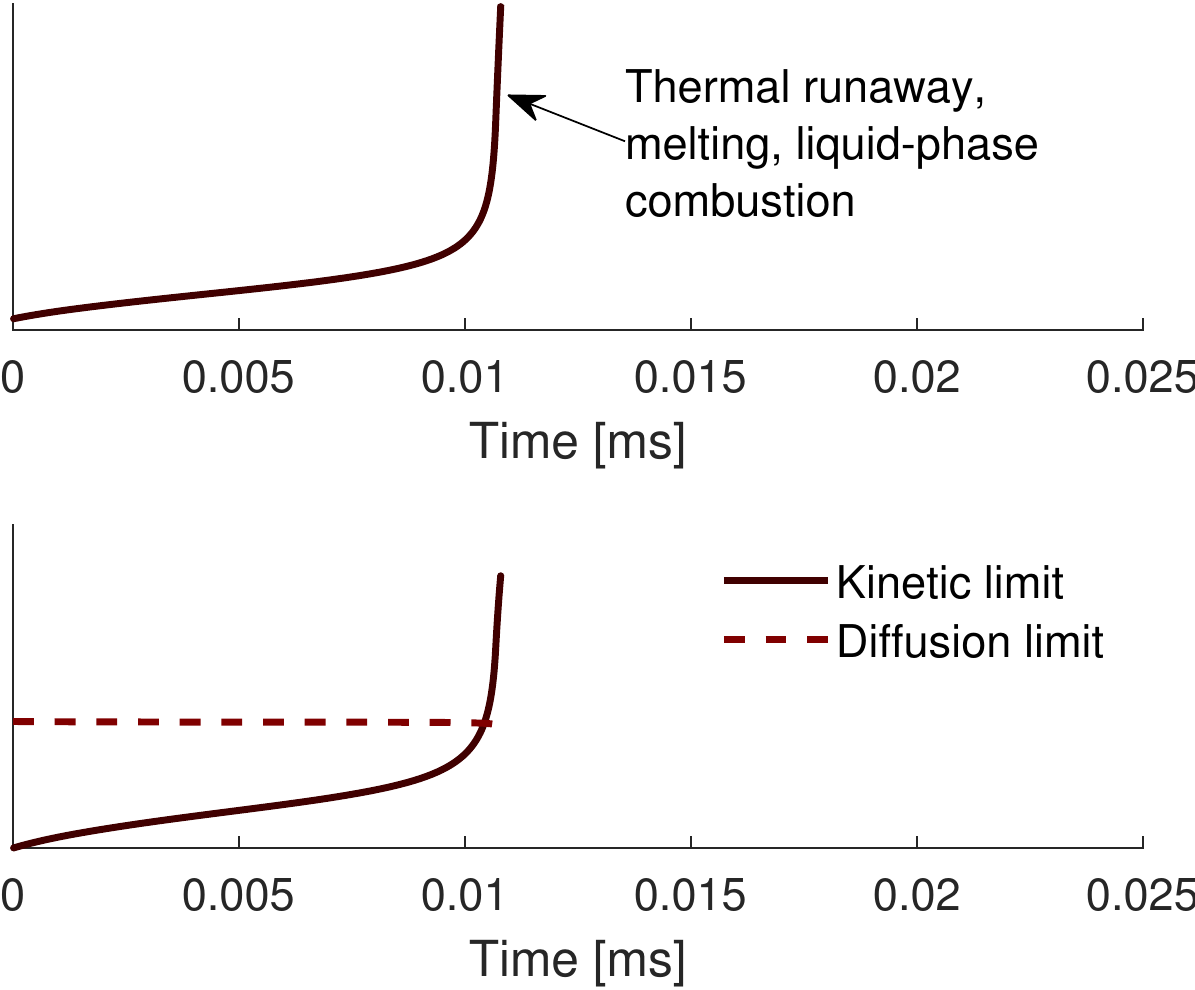}
	\caption{21\% oxidizer concentration.}
	\label{fig:sample_regime_21}
	\end{subfigure}
	
	\caption{Transient particle temperature profile (top row) and oxidizer consumption limits (bottom row) at different bulk gas oxidizer molar fraction for a particle with $\dpzero = 50~\text{nm}$, $X_0 = 10~\text{nm}$, for $\tg = 1022~\text{K} > \tign \approx 1021~\text{K}$.}
	\label{fig:sample_regime_parabolic}
	
\end{figure}

The results presented in the current section are resolved with the implicit method (Section~\ref{sec:balanceknudsen}). The oxidizer molar fraction in the bulk gas $\muox$ is varied and the transient parabolic model is resolved. Since the parabolic kinetics are independent of $\coxp$, $\tign$ remains unaffected by a change in $\muox$. However, Fig.~\ref{fig:sample_regime_parabolic} shows that, for $\tg = 1022~\text{K} > \tign$, the peak particle temperature reached, during its stabilized diffusion-limited combustion after the thermal runaway process, is impacted by $\muox$. For combustion in air ($\muox = 21~\%$, Fig.~\ref{fig:sample_regime_21}), $\tp$ overshoots the melting point of iron and its oxides, the lowest being that of \feo{}--1650~K. For combustion at low oxidizer concentration (e.g. $\muox = 1~\%$, Fig.~\ref{fig:sample_regime_1}), $\tp$ stabilizes at $\approx 1215~\text{K}$ after the thermal runaway process, below the melting point of iron and its oxides. Once the oxide layer grows sufficiently thick, the particle extinguishes; hence, it has burnt out in the solid phase without melting. 

\begin{figure}
	\centering
	
	\begin{subfigure}{0.35\linewidth}
	\centering
	\includegraphics[width=\linewidth]{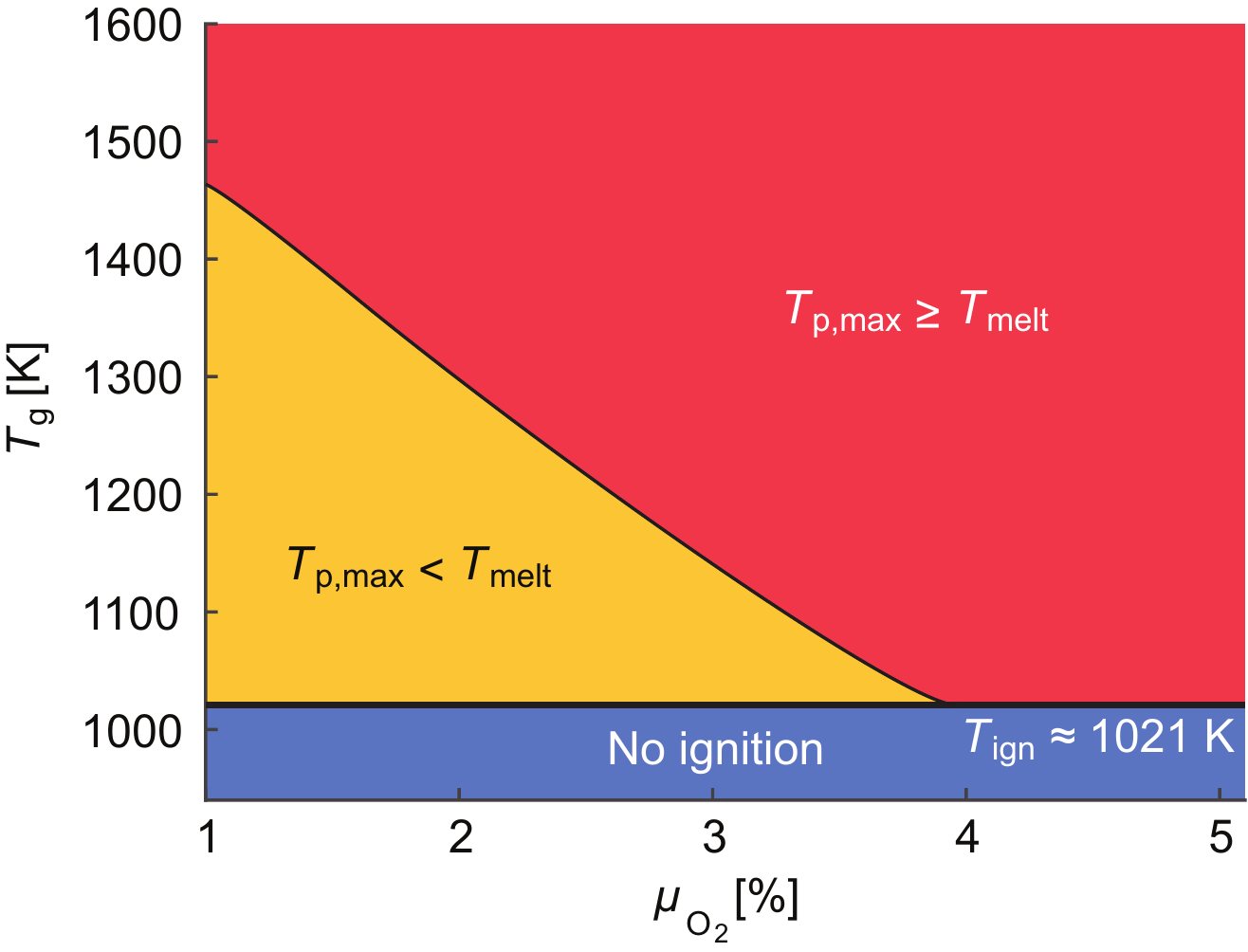}
	\caption{Transition model.}
	\label{fig:regime_parabolic_knu}
	\end{subfigure}
	\begin{subfigure}{0.35\linewidth}
	\centering
	\includegraphics[width=\linewidth]{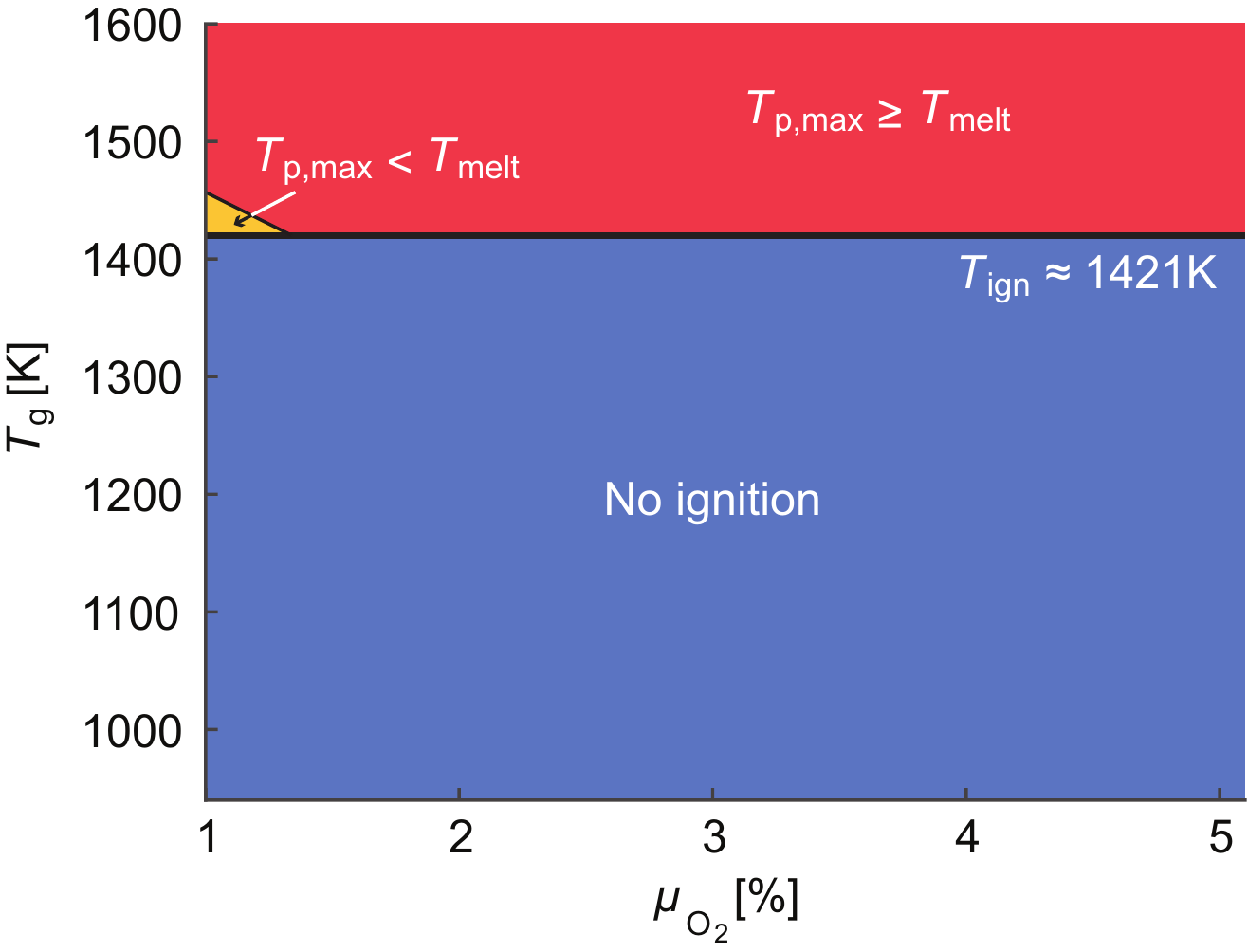}
	\caption{Continuum model.}
	\label{fig:regime_parabolic_co}
	\end{subfigure}
	
	\caption{Burning regime as a function of bulk gas temperature at low bulk gas oxidizer molar fraction, with $\dpzero = 50$~nm, $X_0 = 10$~nm  (parabolic kinetics).}
	\label{fig:regime_parabolic}
	
\end{figure}

The burning regime as a function of $\muox$ and $\tg$ is resolved in Fig.~\ref{fig:regime_parabolic} for $\dpzero = 50$~nm, with the transition and continuum transport models. Since $\tign$ is predicted to be higher with the continuum model (1421~K), the ignited region with $\tpmax < \tmelt$ becomes nearly nonexistent. In comparison, the transition model predicts a much larger region with $\tpmax < \tmelt$, due to the lower $\tign$. The combustion regime where $\tpmax < \tmelt$ is of importance for practical applications: a combustion purely in solid-phase implies no evaporation takes place at the particle surface. This can facilitate collection of iron oxides, which is one of the foundations of the iron fuel economy proposed in Ref.~\cite{bergthorson2018}. This regime is observed at sufficiently low oxidizer concentration, as a function of particle size and bulk gas temperature, as shown in Fig.~\ref{fig:regime_parabolic_knu}. Further experimental and theoretical investigations are required to elucidate this combustion regime.

%\bibliographystyle{../pci}
%\bibliography{../afl_refs}

\subsection{First-order kinetics - results and comparison to parabolic model} \label{sec:results_firstorder}

%%%%%%%%%%%%%%%%%%%%%%%%%%%%%%%%%%%%%%%%%%%
% SAMPLE
%%%%%%%%%%%%%%%%%%%%%%%%%%%%%%%%%%%%%%%%%%%

\subsubsection{Sample results - transient behavior and ignition}

\begin{figure}[h]
	\centering 
	
	\begin{subfigure}{.35\linewidth}
	\centering 
	\includegraphics[width=\linewidth]{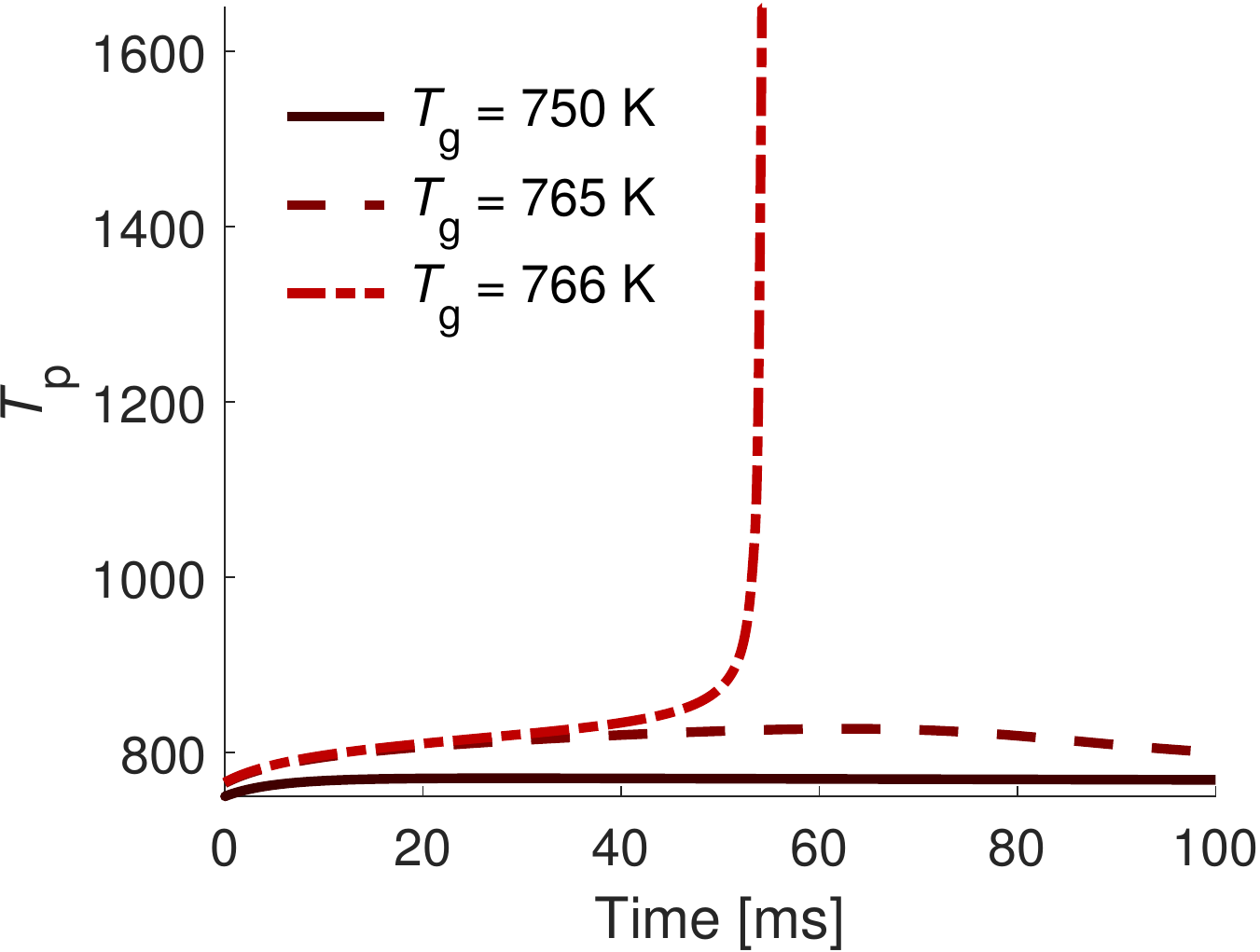}
	\caption{First-order kinetics with $X_0 = 0$~nm. Ignition occurs at $\tg = 766$~K.}
	\label{fig:sampleT_firstorder} 
	\end{subfigure}
	\begin{subfigure}{.35\linewidth}
	\centering 
	\includegraphics[width=\linewidth]{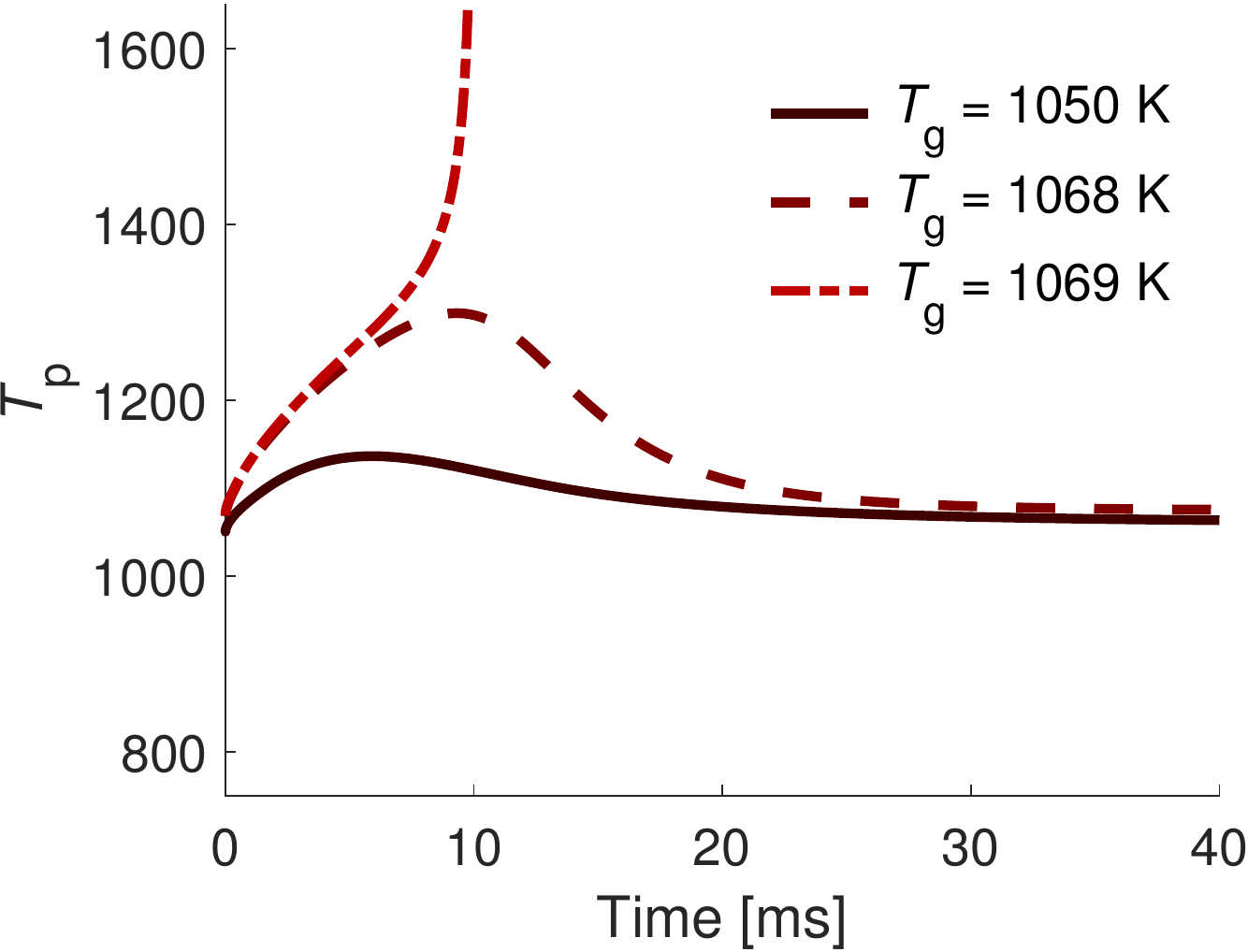}
	\caption{Parabolic kinetics with $X_0 = 1$~nm. Ignition occurs at $\tg = 1069$~K.}
	\label{fig:sampleT_parabolic_1nm} 
	\end{subfigure}
	
	\caption{Particle transient temperature profile for $\dpzero = 20~\upmu$m at different $\tg$.}
	\label{fig:sample_firstorder} 
	
\end{figure}

Figure~\ref{fig:sampleT_firstorder} shows the transient behavior of a particle in different bulk gas temperatures resolved with the first-order kinetic model. The behavior is similar to that observed in Fig.~\ref{fig:sampleT_parabolic_1nm} for the parabolic kinetic model; however, the separation between $\tp$ and $\tg$ is much less pronounced for temperatures below $\tign$ in Fig.~\ref{fig:sampleT_firstorder}. This is explained by the inverse dependence of the parabolic kinetics on the growing oxide thickness, $X$, which can lead to a quenching of the particle after the thermal runaway process has begun and can cause $\Delta T_\text{max} = \tpmax - \tg$ to reach high values for $\tg < \tign$. In contrast, the first-order kinetics show no dependency on $X$; hence, the ignition phenomenon exhibits higher criticality. An additional factor which contributes to the higher $\Delta T_\text{max}$ in the parabolic kinetics is the higher activation temperature of the reactions (see Table~\ref{tab:properties}).

%%%%%%%%%%%%%%%%%%%%%%%%%%%%%%%%%%%%%%%%%%%
% DP0
%%%%%%%%%%%%%%%%%%%%%%%%%%%%%%%%%%%%%%%%%%%

\subsubsection{Ignition behavior - effect of particle size}

\begin{figure}
    \centering
    
    \begin{subfigure}{0.35\linewidth}
    \centering
    \includegraphics[width = \linewidth]{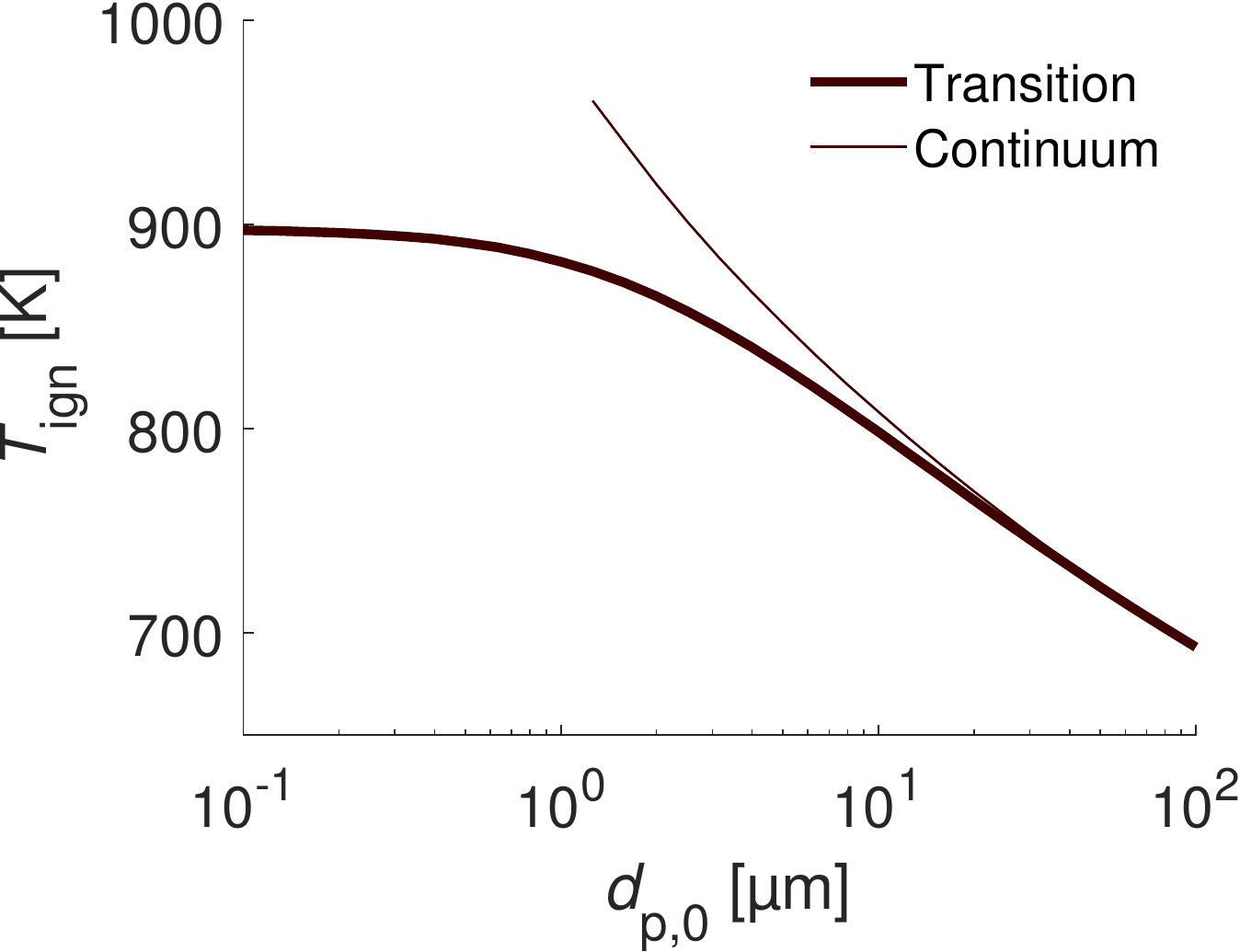}
    \caption{First-order kinetics with $X_0 = 0$~nm.}
    \label{fig:ignition_firstorder}
    \end{subfigure}
    \begin{subfigure}{0.35\linewidth}
    \centering
    \includegraphics[width = \linewidth]{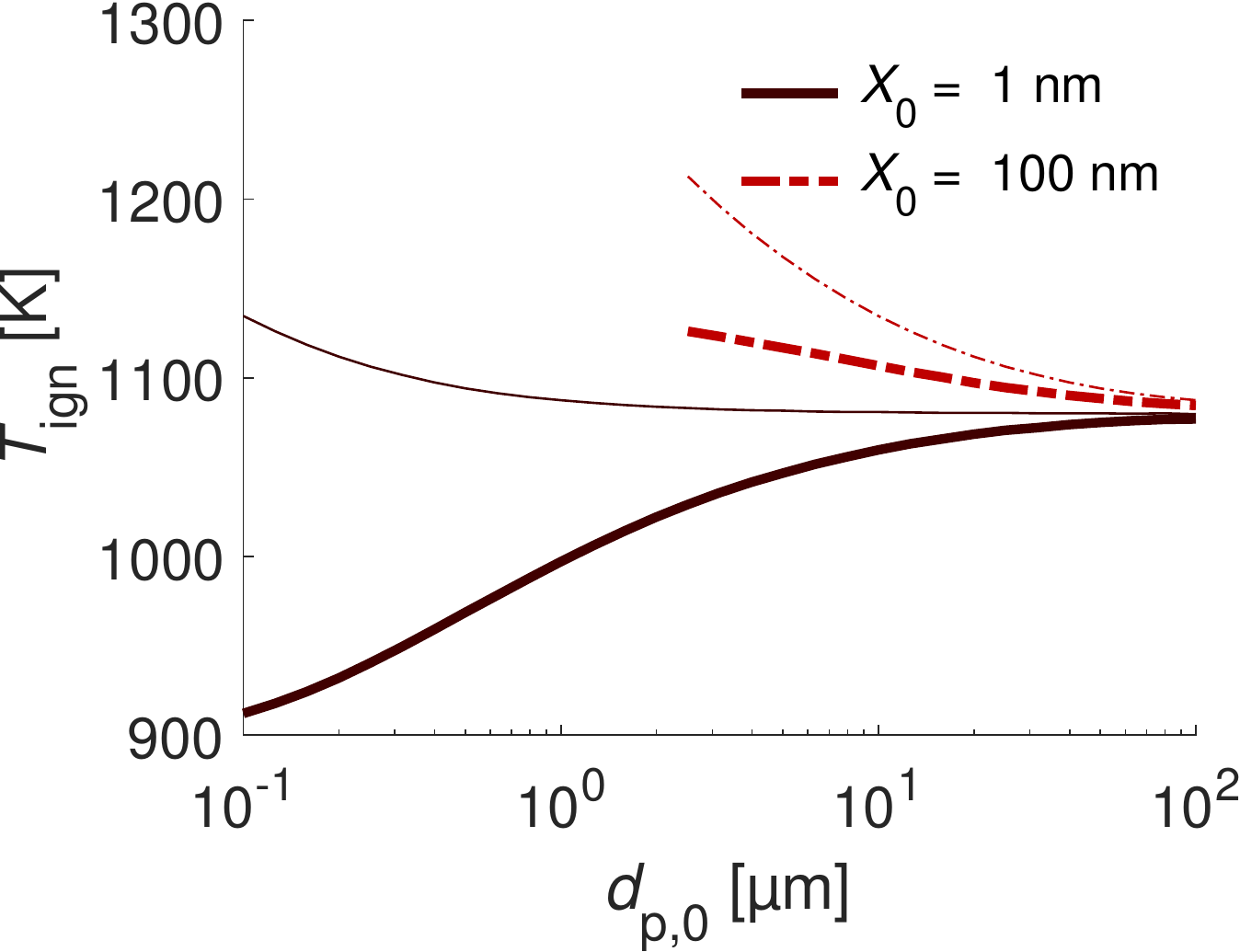}
    \caption{Parabolic kinetics for different values of $X_0$.}
    \label{fig:ignition_parabolic}
    \end{subfigure}
    
    \caption{Ignition temperature as a function of initial particle size resolved with Knudsen transition (thick lower branches) and continuum (thin upper branches) transport models, with the unsteady analysis.}
\end{figure}

Figure~\ref{fig:ignition_firstorder} shows  $\tign$ vs $\dpzero$ with the first-order kinetic model, with the continuum and transition transport methods. The transition model predicts an increase of $\tign$ with decreasing $\dpzero$, which is in contrast with the parabolic kinetics shown in Fig.~\ref{fig:ignition_parabolic}, which may predict a decrease of $\tign$ with decreasing $\dpzero$. In fact, the first-order kinetics exhibit a dependence on $\coxp$. Consequently, the decaying \ox{} molecule-particle collision rate with decreasing particle size leads to a decrease of the kinetic rate for smaller particles. This competes with the thermal insulating effect of the transition heat transport regime, and the net result is an increase of $\tign$ with decreasing $\dpzero$. 

An additional result from Fig.~\ref{fig:ignition_firstorder} is a plateau of $\tign$ for small particles and a removal of the small-particle ignition degeneration limit, which was demonstrated by Soo et al. \cite{soo2018} for first-order oxidation kinetics. In fact, under continuum transport analysis, particles with an initial diameter below a critical particle size become incapable of accumulating sufficient heat to undergo thermal runaway. This ignition degeneration limit is also observed in the current work, as shown from the continuum line in Fig.~\ref{fig:ignition_firstorder}. However, the transition transport model shows this degeneration limit is removed. 

The plateau of $\tign$ for small particles can be further elucidated through a mathematical analysis. Following the first Semenov ignition criterion, $\qdot = \qdotr$, Eq.~(\ref{eq:heatloss}) and (\ref{eq:qdotrfirstorder}) can be equated and re-arranged to yield:
\beq \label{eq:semenov1}
\tg = \tp - \f{\D^* \coxg \nu_\f{\feo}{\ox} \qfeo k_1}{k^*} \Bigg( \f{2 \rp^2 + \gt \th \rp + \gt \th^2}{ \big[ 2 \rp^2 + \gm \rp \th + \gm \th^2 \big] k_1 + 2 \D^* (\rp + \th)} \Bigg).
\eeq
The second Semenov ignition criterion is $\d\qdot/\d \tp = \d \qdotr/\d \tp$. Using,
\beq
k_1 = \kinfone \exp \bigg(\f{-\taone}{\tp} \bigg) 
\Rightarrow \f{\d k_1}{\d \tp} = k_1 \f{\taone}{\tp^2}
\eeq
Eq.~(\ref{eq:heatloss}) and (\ref{eq:qdotrfirstorder}) can be differentiated with respect to $\tp$ and equated to yield:
\beq \label{eq:semenov2}
\f{k^*}{2 \rp^2 + \gt \th \rp + \gt \th^2} =  2  {(\D^*)}^2 \coxg \nu_\f{\feo}{\ox} \qfeo k_1 \bigg( \f{\ta}{\tp^2}\bigg)  \f{  \rp + \th}{ \Big \{ \big[  2 \rp^2 + \gm \rp \th + \gm \th^2 \big] k_1 + 2 \D^* (\rp + \th) \Big\}^2}.
\eeq 
In the small particle limit, $\rp \Lt \th$, which allows to reduce Eq.~(\ref{eq:semenov1}) and (\ref{eq:semenov2}) to:
\beqarr
\label{eq:semenovsmall1} \tg &=& \tp - \f{\D^* \coxg \nu_\f{\feo}{\ox} \qfeo k_1}{k^*} \bigg( \f{\gt \th}{\gm \th k_1 + 2 \D^* } \bigg) \\
\label{eq:semenovsmall2} \tp^2 &=& \f{2 {(\D^*})^2 \coxg \nu_\f{\feo}{\ox} \qfeo k_1 \taone}{k^*} \bigg( \f{\gt \th}{ [\gm \th k_1 + 2 \D^* ]^2} \bigg).
\eeqarr
The solution of the system defined by Eq.~(\ref{eq:semenovsmall1}) and (\ref{eq:semenovsmall2}) only depends on: the thermophysical and transport properties of the gas, the TAC and MAC through the geometrical Knudsen heat and mass transfer factors, the reaction kinetic parameters, and the molecular mean free path. Since the system always has a solution and is independent of $\rp$, no small-particle ignition degeneration limit is observed, and a plateau of $\tign$ is observed for small particles. 

\begin{figure}[h]
    \centering
    \includegraphics[width = 0.4\linewidth]{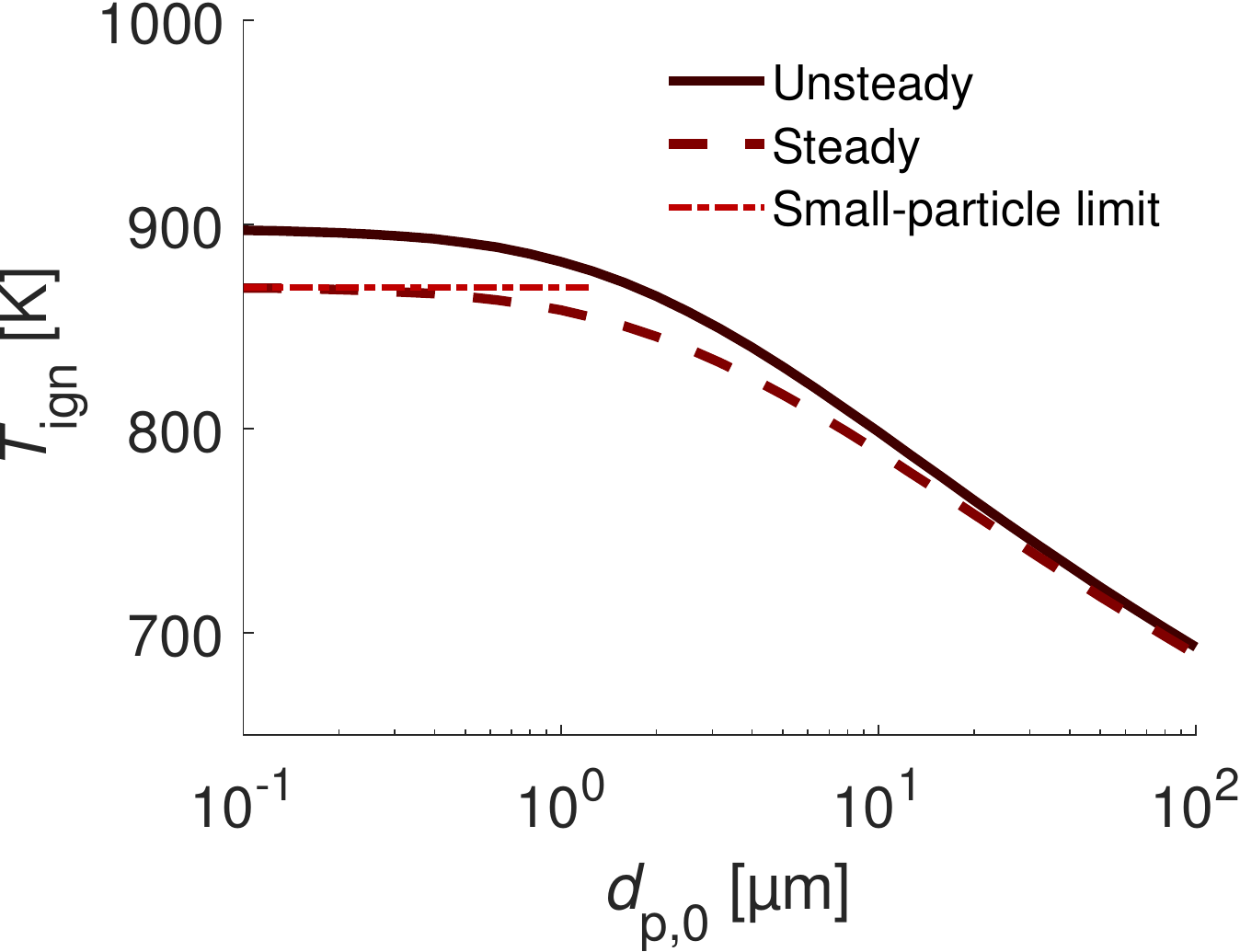}
    \caption{Ignition temperature as resolved by the unsteady and steady analyses (first-order kinetics); comparison to the small-particle ignition plateau calculated through Eq.~(\ref{eq:semenovsmall1}) and (\ref{eq:semenovsmall2}).}
	\label{fig:steady_firstorder}
\end{figure}

The system defined by Eq.~(\ref{eq:semenovsmall1}) and (\ref{eq:semenovsmall2}) is solved numerically, and the result is compared to the unsteady and steady ignition models for the first-order kinetics; results are shown in Fig.~\ref{fig:steady_firstorder}. The steady analysis tends towards the  plateau of $\tign$ calculated through Eq.~(\ref{eq:semenovsmall1}) and (\ref{eq:semenovsmall2}), which is $\approx$ 869~K. This plateau is below the observed plateau of $\approx$ 900~K in the unsteady analysis. In fact, as is the case with the parabolic model, the steady analysis under-predicts $\tign$, since the growth of the oxide is neglected, while it represents inert thermal mass to be heated during the thermal runaway of the particle. However, in contrast to the parabolic kinetics, the difference between the two models is negligible, since the growth of the oxide layer has no adverse effect on the first-order kinetic rate of oxidation.

%\bibliographystyle{../pci}
%\bibliography{../afl_refs}

%\subfile{sections/4.1-parabolic.tex}
%\subfile{sections/4.2-firstorder.tex}
%\bibliographystyle{../pci}
%\bibliography{../afl_refs}

%\subfile{sections/4.1-parabolic.tex}
%\subfile{sections/4.2-firstorder.tex}

\section{Discussion} \label{sec:discussion}

%%%%%%%%%%%%%%%%%%%%%%%%%%%%%%%%%%%%%%%%%%%
% KINETICS
%%%%%%%%%%%%%%%%%%%%%%%%%%%%%%%%%%%%%%%%%%%

\subsection{Oxidation kinetics of iron particles}

\begin{figure}
    \centering
    \includegraphics[width = 0.25\linewidth]{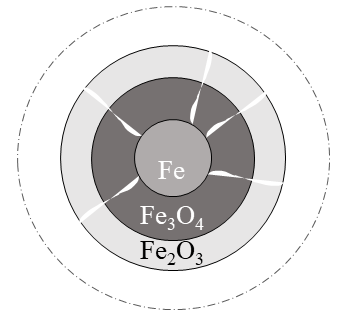}
    \caption{Oxidation model with the formation of cracks (direct oxidizer diffusion channels) in the oxide layer.}
    \label{fig:oxidation_tga}
\end{figure}

The parabolic kinetic model used in the current study is based on the model developed by Mi et al. \cite{mi2022}, who calibrated their kinetic parameters to the experimental work of Pa\"{i}dassi \cite{paidassi1958}. In Ref. \cite{paidassi1958}, the kinetics of iron were studied through the isothermal growth rate of iron oxides on the surface of iron films in the temperature range 973--1523~K. One possible limitation of the work carried by Mi et al. \cite{mi2022} is that the experimental results of Pa\"{i}dassi were based on bulk material--iron films--and were used to predict the kinetics of iron particles. However, the mechanisms of lattice diffusion may not accurately describe particle kinetics. In fact, Lysenko et al. \cite{lysenko2014} studied the kinetics of iron directly with particles in their TGA studies. In Ref. \cite{lysenko2014}, an activation energy of 110~kJ/mol for the formation of \ftt{} was reported, the highest iron oxide. In comparison, Pa\"{i}dassi \cite{paidassi1958} reported an activation energy of 169~kJ/mol, while the survey conducted by Chen and Yeun \cite{chen2003} resulted in an activation energy of 155~kJ/mol. Hence, the noticeable difference in activation energy may be attributed to mechanisms intrinsic to the sample geometry, such as the possible formation of small cracks across the oxide layers, which creates direct diffusion channels for the oxygen to react with the iron core, as depicted in Fig.~\ref{fig:oxidation_tga}. Further experimental endeavors are required to elucidate this question.
 
An additional possible limitation of the parabolic kinetic model used is that the current model extends the Pa\"{i}dassi \cite{paidassi1958} kinetics to $\approx 900$~K in the unsteady analysis (Fig.~\ref{fig:steady_parabolic}). However, the kinetics are calibrated in the range 973--1573~K. To more accurately predict $\tign$ for particles $\lesssim~1~\upmu$m, improved kinetics for $\tp <$ 973~K should be used in future studies.

%%%%%%%%%%%%%%%%%%%%%%%%%%%%%%%%%%%%%%%%%%%
% KNUDSEN LAYER
%%%%%%%%%%%%%%%%%%%%%%%%%%%%%%%%%%%%%%%%%%%

\subsection{Formulation of the Knudsen layer thickness}

The primary difficulty of the boundary sphere method resides in determining the appropriate Knudsen layer thickness $\th$ for the flux-matching conditions. The perhaps most widely applied formulation was provided by Wright \cite{wright1960} in a derivation based on the Knudsen cosine law, which takes into account particle curvature effects to calculate the effective average free-molecular transport region around the particle surface. However, the exact formulation of $\th$ was shown to yield only marginal variations of the flux rates in the boundary sphere method, provided that it is a factor close to unity of the mean free path \cite{liu2006, fuchs1959}. In the current work, the thickness of the Knudsen layer, $\th$, is set equal to the molecular mean free path in the bulk gas, which allows a constant value of $\th$ to be used for the transient analyses. This approximation is justified by its minimal impact on the transition heat and mass transport rates.

%\bibliographystyle{../pci}
%\bibliography{../afl_refs}

\section{Conclusions} \label{sec:conclusion}

The current study quantitatively assesses Knudsen transition transport effects on the ignition behavior of fine iron particles. A computational model considering two possible high-temperature solid-phase iron oxidation models--parabolic and first-order kinetics--and coupling them to a boundary-sphere flux-matching transition transport method was implemented. The model was solved to resolve the critical gas temperature at which particle ignition can occur as a function of particle size, and the results predicted by the two kinetic models were compared. Additionally, results obtained with the transition transport analysis were compared to a continuum transport approach for both kinetic models.

In the parabolic kinetic model, the ignition temperature was shown to behave non-monotonically with decreasing particle size. Transition transport processes lead to a thermal insulating effect with decreasing particle size, which facilitates particle ignition. However, the increasing oxide layer proportion with decreasing particle size adversely affects the reaction kinetics, impeding particle ignition. These two effects compete and the reaction model should be solved unsteadily to assess the overall impact on the ignition temperature as a function of particle size and initial oxide layer thickness. The unsteady analysis is increasingly important for larger particles and smaller initial oxide layers. Continuum transport modeling was shown to predict ignition temperatures to within 10--30~K or less of the transition transport model for particles with an initial diameter exceeding 30~$\upmu$m. The 30~$\upmu$m limit agrees with the order of magnitude reported by a majority of previous researchers for the onset of transition effects. In the small-particle limit, the transition transport analysis revealed a combustion regime where particles ignite and burn in the diffusion-limited regime below the melting point of iron and its oxides.

In the first-order kinetic model, the ignition temperature was shown to monotonically decrease with increasing particle size. The continuum small-particle ignition degeneration limit was shown to be removed with the transition transport analysis, due to the independence of ignition temperature on the particle size in the small-particle limit. The ignition temperature was shown to tend towards a plateau in the small-particle limit. The different qualitative trends between the parabolic and the first-order kinetic models provide a comparison basis for future experimental work that aims to validate the high-temperature solid-phase oxidation and ignition behavior of single iron particles.

%\bibliographystyle{../pci}
%\bibliography{../afl_refs}

\appendix
\renewcommand{\theequation}{\thesection.\arabic{equation}}

\setcounter{equation}{0}

\section{Explicit mass transfer factor in the boundary sphere flux-matching method} \label{sec:a1}

The detailed derivation of Eq.~(\ref{eq:betam}) is provided in the current Appendix. Liu et al. \cite{liu2006} derived an explicit formulation of the heat transport rate between a spherical particle and a gas in the boundary sphere flux-matching method,
\beq \label{eqapp:betat}
\betat = \f{\qdot}{\qdotc} = \bigg(\f{1}{1+\Kn} + \f{1}{2} \gt \Kn \bigg)^{-1}
\eeq
where $\qdot$ is the actual heat loss rate from the particle accounting for transition transport effects; $\qdotc = 4 \pi \rp k (\tp - \tg)$ is the heat loss formulated in the continuum approximation, with $\rp$ the particle radius, $k$ the gas mixture-averaged thermal conductivity, and $\tp$, $\tg$ respectively the particle and bulk gas temperature; $\Kn = \mfp/\rp$ is the Knudsen number based on the bulk gas mean free path $\mfp$; and,
\beq \label{eqapp:gt}
\gt = \f{8f}{\alphat (\gamma+1)}
\eeq
is the geometry-dependent heat transfer factor, with $f = (9\gamma - 5)/4$ the Eucken factor, $\gamma$ the heat capacity ratio, and $\alphat$ the thermal accommodation coefficient (TAC). Equation~(\ref{eqapp:betat}) incorporates the assumption that the Knudsen layer thickness is formulated as the mean free path of the bulk gas, $\theta = \mfp$. As well, Eq.~(\ref{eqapp:betat}) implies the ratio of the mean thermal molecular speeds in the bulk gas and at the Knudsen layer surface tends to 1:
\beq \label{eqapp:speedratio1}
\f{\cbarg}{\cbarth} \equiv \bigg(\f{\tg}{\tth}\bigg)^{\nf{1}{2}} \bigg(\f{\mth}{\mg}\bigg)^{\nf{1}{2}} \approx 1.
\eeq
In Eq.~(\ref{eqapp:speedratio1}), $\mth$ and $\mg$ are respectively the average individual molecular mass of the gas molecules at the Knudsen layer surface and in the bulk gas. This ratio tends to 1 if the concentrations of the gaseous species at the limiting sphere surface and in the bulk gas are similar. As well, $(\mth/\mg)^{\nf{1}{2}} \rightarrow 1$ if the inert gas and the oxidizer have a similar molecular mass, independent of the concentration gradients. Such is the case for a gaseous mix of \ox{} and \inert{}. Equation~(\ref{eqapp:betat}) as well assumes $(\tg/\tth)^{\nf{1}{2}} \rightarrow 1$, in other words there is a small temperature difference between the Knudsen layer surface and the bulk gas. This is always the case in the pre-ignition phase of an iron particle.

Equation~(\ref{eqapp:betat}) incorporates the methodology of the boundary sphere method, while allowing to compute the heat transfer rate without solving a coupled system of nonlinear equations. A similar term $\betam$ can be derived for the boundary sphere mass transport rate, following the same approach as in \cite{liu2006}. The oxidizer transport inside and outside the Knudsen layer is equivalently described by:
\beqarr 
\label{eqapp:particle} \mdot &=& \alpham \pi \rp^2 (\cth \cbarth - \cp \cbarp) \\
\label{eqapp:layer} \mdot &=& 4 \pi (\rp + \th) \D (\cg - \cth).
\eeqarr
where $\D$ is the oxidizer mass diffusivity in the gas mixture, $\alpham$ is the mass accommodation coefficient (MAC), $C$ is the oxidizer concentration, and the subscripts g, $\th$, and p respectively denote in the bulk gas, at the Knudsen layer surface, and at the particle surface. The symbol $C$ is used to describe oxidizer concentration instead of $\cox$ as in previous sections to lighten the notation. If the mass transport would occur completely in the free-molecular or the continuum regime from the bulk gas to the particle surface, it would respectively be described by:
\beqarr
 \label{eqapp:fm} \mdotfm &=& \alpham \pi \rp^2 (\cg \cbarg - \cp \cbarp) \\
 \label{eqapp:cont} \mdotc &=& 4 \pi \rp \D (\cg - \cp).
\eeqarr
Multiplying Eq.~(\ref{eqapp:particle}) by $\mdotc/\mdotfm$ results in:
\beq
\mdot \f{\mdotc}{\mdotfm} = \alpham \pi \rp^2 (\cth \cbarth - \cp \cbarp) \f{4 \pi \rp \D (\cg - \cp)}{\alpham \pi \rp^2 (\cg \cbarg - \cp \cbarp)} = 4 \pi \rp \D \bigg(\f{\cth \cbarth - \cp \cbarp}{\cbarg}\bigg) \f{\cg - \cp}{ \cg - \cp (\cbarp/\cbarg)}.
\eeq
Using the assumption $\cbarth/\cbarg \approx 1$ and introducing the assumption $\cbarp/\cbarg \approx 1$, this simplifies to:
\beq \label{eqapp:add1}
\mdot \f{\mdotc}{\mdotfm} = 4 \pi \rp \D (\cth - \cp).
\eeq
Now the ratio of the transport rates can be expanded to,
\beq \label{eqapp:use}
\f{\mdotc}{\mdotfm} = \f{4 \pi \rp \D (\cg - \cp)}{\alpham \pi \rp^2 (\cg \cbarg - \cp \cbarp)} = \f{4 \D (\cg - \cp)}{\alpham \rp \cbarg (\cg - \cp (\cbarp/\cbarg))} = \f{4 \D}{\alpham \rp \cbarg}
\eeq
where the assumption $\cbarp/\cbarg \approx 1$ was used. Additionally, Eq.~(\ref{eqapp:layer}) can be re-arranged to:
\beq \label{eqapp:add2}
\mdot \f{\rp}{\rp + \theta}  = 4 \pi \rp \D (\cg - \cth).
\eeq
Adding Eq.~(\ref{eqapp:add1}) and~(\ref{eqapp:add2}) and using Eq.~(\ref{eqapp:use}) yields:
\beq
\mdot \bigg( \f{\rp}{\rp + \theta} +  \f{4 \D}{\alpham \rp \cbarg} \bigg) = 4 \pi \rp \D (\cth - \cp) + 4 \pi \rp \D (\cg - \cth) \equiv \mdotc.
\eeq
Using the approximation $\th = \mfp$, the definition of the Knudsen number $\Kn = \mfp/\rp$, and re-arranging yields:
\beq \label{eqapp:expand}
\f{\mdot}{\mdotc} = \bigg(\f{1}{1+\Kn} + \f{4\D}{\alpham \cbarg \mfp} \Kn \bigg)^{-1}.
\eeq
Using the definition of the mean free path \cite{liu2006}:
\beq
\mfp = \f{k (\gamma - 1)}{f p} \bigg( \f{\pi \mg \tg}{2 \KB}\bigg)^{\nf{1}{2}}
\eeq
where $p$ is the bulk gas pressure, and the result $[\pi \mg \tg/(2 \KB)]^{\nf{1}{2}} \equiv 2 \tg/\cbarg$, the factor in the second term on the right-hand-side of Eq.~(\ref{eqapp:expand}) can be re-written as,
\beq \label{eqapp:subs}
\f{4 \D}{\alpham \cbarg \mfp} = \f{4 \D}{\alpham \cbarg} \f{f p}{k (\gamma-1)} \f{\cbarg}{2 \tg} = \f{2 \D f p}{\alpham k (\gamma-1) \tg} \equiv \f{2 f}{\alpham (\gamma-1)} \f{\D \rho \RG}{k}
\eeq
where the ideal gas law $p = \rho \RG \tg$ was used, with $\rho$ the gas density and $\RG$ the individual gas constant. Now the definition of the Lewis number yields,
\beq
\Le = \f{\alpha}{\D} = \f{k}{\rho c_p \D} \Rightarrow \f{\D \rho}{k} = \f{1}{c_p \Le}
\eeq
where $\alpha$ is the thermal diffusivity and $c_p$ is the heat capacity at constant pressure. Substituting in Eq.~(\ref{eqapp:subs}):
\beqarr
\nonumber \f{4 \D}{\alpham \cbarg \mfp} &=& \f{2f}{\alpham (\gamma - 1)} \f{\RG}{c_p \Le} \\
\nonumber  \because c_p = c_v + \RG &\Rightarrow& 1 = \f{1}{\gamma} + \f{\RG}{c_p} \Rightarrow \f{\RG}{c_p} = \f{\gamma-1}{\gamma} \\
\label{eqapp:uselast} \therefore \f{4 \D}{\alpham \cbarg \mfp} &=& \f{2 f}{\alpham \gamma \Le }.
\eeqarr
Following Liu et al.'s \cite{liu2006} appellation of $\gt$ as the geometry-dependent heat transfer factor, the new geometry-dependent mass transfer factor is defined as,
\beq \label{eqapp:gn}
\gm = \f{4f}{\alpham \gamma \Le}
\eeq
which is a non-dimensional number. Substituting Eq.~(\ref{eqapp:uselast}) in Eq.~(\ref{eqapp:expand}) and using the above-defined $\gm$ results in,
\beq \label{eqapp:betam}
\betam = \f{\mdot}{\mdotc} = \bigg(\f{1}{1+\Kn} + \f{1}{2} \gm \Kn \bigg)^{-1}
\eeq
which has a form identical to Eq.~(\ref{eqapp:betat}). The new boundary sphere mass transfer factor $\betam$ provided by Eq.~(\ref{eqapp:betam}) allows to compute the oxidizer mass transport rate from the bulk gas to the particle surface explicitly, by applying a transitional correction factor to the continuum rate. The heat and mass transport can therefore both be resolved explicitly with the boundary sphere method. Eq.~(\ref{eqapp:betat}) and~(\ref{eqapp:betam}) are valid for arbitrary Knudsen number, and incorporate the assumptions $\cbarg/\cbarp \approx 1$, $\cbarg/\cbarth \approx 1$, $\theta = \mfp$. 

Using Eq.~(\ref{eqapp:gt}) and (\ref{eqapp:gn}), the heat and mass transfer factors are related by:
\beq
\f{\gt}{\gm} = \f{8f}{\alphat (\gamma+1)} \f{\alpham \gamma \Le}{4 f} \Rightarrow \f{\gt}{\gm} = 2 \f{\alpham}{\alphat}\bigg(\f{\gamma}{\gamma+1}\bigg) \Le.
\eeq

%\bibliographystyle{../pci}
%\bibliography{../afl_refs}

%\subfile{sections/A2-method.tex}

%%%%%%%%%%%%%%%%%%%%%%%%%%%%%%%%%%%%%%%%%%%
% ACKNOWLEDGEMENTS
%%%%%%%%%%%%%%%%%%%%%%%%%%%%%%%%%%%%%%%%%%%

\section*{Acknowledgements}

The authors thank the members of the Alternative Fuels Laboratory of McGill University for useful discussions in developing this paper. This project is undertaken with the financial support of the Canadian Space Agency (CSA), the Fonds de Recherche due Québec (FRQ), and the Natural Sciences and Engineering Research Council of Canada (NSERC).

%%%%%%%%%%%%%%%%%%%%%%%%%%%%%%%%%%%%%%%%%%%
% REFERENCES
%%%%%%%%%%%%%%%%%%%%%%%%%%%%%%%%%%%%%%%%%%%

%\bibliographystyle{cas-model2-names}
\bibliographystyle{pci}
\bibliography{afl_refs}

\begin{thebibliography}{10}
\expandafter\ifx\csname url\endcsname\relax
  \def\url#1{\texttt{#1}}\fi
\expandafter\ifx\csname urlprefix\endcsname\relax\def\urlprefix{URL }\fi
\expandafter\ifx\csname href\endcsname\relax
  \def\href#1#2{#2} \def\path#1{#1}\fi

\bibitem{bergthorson2015}
J.~Bergthorson, S.~Goroshin, M.~Soo, P.~Julien, J.~Palecka, D.~Frost,
  D.~Jarvis, Direct combustion of recyclable metal fuels for zero-carbon heat
  and power, Applied Energy 160 (2015) 368--382.

\bibitem{bergthorson2018}
J.~M. Bergthorson, Recyclable metal fuels for clean and compact zero-carbon
  power, Progress in Energy and Combustion Science 68 (2018) 169--196.

\bibitem{frank1955}
D.~A. Frank-Kamenetskii, Diffusion and heat exchange in chemical kinetics,
  Princeton University Press, 1955.

\bibitem{soo2018}
M.~Soo, X.~Mi, S.~Goroshin, A.~J. Higgins, J.~M. Bergthorson, Combustion of
  particles, agglomerates, and suspensions--a basic thermophysical analysis,
  Combustion and Flame 192 (2018) 384--400.

\bibitem{gopalakrishnan2011}
R.~Gopalakrishnan, T.~Thajudeen, C.~J. Hogan~Jr, Collision limited reaction
  rates for arbitrarily shaped particles across the entire diffusive knudsen
  number range, The Journal of chemical physics 135~(5) (2011) 054302.

\bibitem{kodas1990}
T.~T. Kodas, P.~B. Comita, The role of mass transport in laser-induced
  chemistry, Accounts of Chemical Research 23~(6) (1990) 188--194.

\bibitem{liu2006}
F.~Liu, K.~Daun, D.~R. Snelling, G.~J. Smallwood, Heat conduction from a
  spherical nano-particle: status of modeling heat conduction in laser-induced
  incandescence, Applied physics B 83~(3) (2006) 355--382.

\bibitem{shpara2020}
A.~Shpara, D.~Yagodnikov, A.~Sukhov, Effect of particle size on boron
  combustion in air, Combustion, Explosion, and Shock Waves 56~(4) (2020)
  471--478.

\bibitem{zou2020}
X.~Zou, N.~Wang, L.~Liao, Q.~Chu, B.~Shi, Prediction of nano/micro aluminum
  particles ignition in oxygen atmosphere, Fuel 266 (2020) 116952.

\bibitem{mohan2008}
S.~Mohan, M.~A. Trunov, E.~L. Dreizin, Heating and ignition of metal particles
  in the transition heat transfer regime (2008).

\bibitem{ermoline2018}
A.~Ermoline, Thermal theory of aluminum particle ignition in continuum,
  free-molecular, and transition heat transfer regimes, Journal of Applied
  Physics 124~(5) (2018) 054301.

\bibitem{senyurt2022}
E.~I. Senyurt, E.~L. Dreizin, At what ambient temperature can thermal runaway
  of a burning metal particle occur?, Combustion and Flame 236 (2022) 111800.

\bibitem{mi2022}
X.~Mi, A.~Fujinawa, J.~M. Bergthorson, A quantitative analysis of the ignition
  characteristics of fine iron particles, Combustion and Flame 240 (2022)
  112011.

\bibitem{paidassi1958}
J.~Pa{\"\i}dassi, Sur la cinetique de l'oxydation du fer dans l'air dans
  l'intervalle 700--1250° c, Acta Metallurgica 6~(3) (1958) 184--194.

\bibitem{goursat1973}
A.~G. Goursat, W.~Smeltzer, Kinetics and morphological development of the oxide
  scale on iron at high temperatures in oxygen at low pressure, Oxidation of
  Metals 6~(2) (1973) 101--116.

\bibitem{fuchs1959}
N.~A. Fuchs, Evaporation and droplet growth in gaseous media, Elsevier, 1959.

\bibitem{hazenberg2021}
T.~Hazenberg, J.~van Oijen, Structures and burning velocities of flames in iron
  aerosols, Proceedings of the Combustion Institute 38~(3) (2021) 4383--4390.

\bibitem{chen2003}
R.~Chen, W.~Yeun, Review of the high-temperature oxidation of iron and carbon
  steels in air or oxygen, Oxidation of metals 59~(5) (2003) 433--468.

\bibitem{lysenko2014}
E.~Lysenko, A.~Surzhikov, S.~Zhuravkov, V.~Vlasov, A.~Pustovalov,
  N.~Yavorovsky, The oxidation kinetics study of ultrafine iron powders by
  thermogravimetric analysis, Journal of Thermal Analysis and Calorimetry
  115~(2) (2014) 1447--1452.

\bibitem{xu2000}
C.~Xu, W.~Gao, Pilling-bedworth ratio for oxidation of alloys, Material
  Research Innovations 3~(4) (2000) 231--235.

\bibitem{kennard1938}
E.~H. Kennard, Kinetic Theory of Gases, With an Introduction to Statistical
  Mechanics, McGraw-Hill, New York, 1938.

\bibitem{qu2001}
X.~Qu, E.~Davis, B.~Swanson, Non-isothermal droplet evaporation and
  condensation in the near-continuum regime, Journal of aerosol science 32~(11)
  (2001) 1315--1339.

\bibitem{goodman1976}
F.~O. Goodman, H.~Y. Wachman, Dynamics of gas-surface scattering, Elsevier,
  1976.

\bibitem{wagner1982}
P.~E. Wagner, Aerosol growth by condensation, in: Aerosol Microphysics II,
  Springer, 1982, pp. 129--178.

\bibitem{wright1960}
P.~Wright, On the discontinuity involved in diffusion across an interface (the
  $\delta$ of fuchs), Discussions of the Faraday Society 30 (1960) 100--112.

\bibitem{saxena1989}
S.~C. Saxena, R.~K. Joshi, Thermal accommodation and adsorption coefficients of
  gases (1989).

\bibitem{king1978}
D.~A. King, Kinetics of adsorption, desorption, and migration at singlecrystal
  metal surfaces, Critical Reviews in Solid State and Material Sciences 7~(3)
  (1978) 167--208.

\bibitem{barker1984}
J.~A. Barker, D.~J. Auerbach, Gas—surface interactions and dynamics; thermal
  energy atomic and molecular beam studies, Surface Science Reports 4~(1-2)
  (1984) 1--99.

\bibitem{shin1965}
H.~Shin, On the effect of adsorbed particles on the accommodation coefficients,
  The Journal of Chemical Physics 42~(10) (1965) 3442--3445.

\bibitem{sipkens2018}
T.~Sipkens, K.~Daun, Effect of surface interatomic potential on thermal
  accommodation coefficients derived from molecular dynamics, The Journal of
  Physical Chemistry C 122~(35) (2018) 20431--20443.

\bibitem{song1987}
S.~Song, M.~Yovanovich, Correlation of thermal accommodation coefficient for
  engineering surfaces, ASME HTD 69 (1987) 107--116.

\bibitem{glasstone1941}
S.~Glasstone, K.~J. Laidler, H.~Eyring, The theory of rate processes; the
  kinetics of chemical reactions, viscosity, diffusion and electrochemical
  phenomena, Tech. rep., McGraw-Hill Book Company, (1941).

\bibitem{chase1991}
M.~Chase, NIST-JANAF Thermochemical Tables, 4th Edition, American Institute of
  Physics, -1, 1991.

\bibitem{hubbard1975}
G.~Hubbard, V.~Denny, A.~Mills, Droplet evaporation: effects of transients and
  variable properties, International journal of heat and mass transfer 18~(9)
  (1975) 1003--1008.

\bibitem{mcbride1993}
B.~J. McBride, Coefficients for calculating thermodynamic and transport
  properties of individual species, Vol. 4513, NASA Langley Research Center,
  1993.

\bibitem{fuller1966}
E.~N. Fuller, P.~D. Schettler, J.~C. Giddings, New method for prediction of
  binary gas-phase diffusion coefficients, Industrial \& Engineering Chemistry
  58~(5) (1966) 18--27.

\bibitem{mikami1966}
H.~Mikami, Y.~Endo, Y.~Takashima, Heat transfer from a sphere to rarefied gas
  mixtures, International Journal of Heat and Mass Transfer 9~(12) (1966)
  1435--1448.

\bibitem{springer1965}
G.~S. Springer, S.~W. Tsai, Method for calculating heat conduction from spheres
  in rarefied gases, The Physics of Fluids 8~(8) (1965) 1561--1563.

\bibitem{daun2013}
K.~Daun, T.~Sipkens, J.~Titantah, M.~Karttunen, Thermal accommodation
  coefficients for laser-induced incandescence sizing of metal nanoparticles in
  monatomic gases, Applied Physics B 112~(3) (2013) 409--420.

\bibitem{sipkens2014}
T.~Sipkens, N.~Singh, K.~Daun, N.~Bizmark, M.~Ioannidis, J.~Titantah,
  M.~Karttunen, Time resolved laser induced incandescence for sizing
  aerosolized iron nanoparticles, in: ASME International Mechanical Engineering
  Congress and Exposition, Vol. 46569, American Society of Mechanical
  Engineers, 2014, p. V08BT10A051.

\bibitem{sipkens2015}
T.~Sipkens, N.~Singh, K.~Daun, N.~Bizmark, M.~Ioannidis, Examination of the
  thermal accommodation coefficient used in the sizing of iron nanoparticles by
  time-resolved laser-induced incandescence, Applied Physics B 119~(4) (2015)
  561--575.

\bibitem{sipkens2015b}
T.~Sipkens, K.~Daun, J.~Titantah, M.~Karttunen, Quantifying the thermal
  accommodation coefficient for iron surfaces using molecular dynamics
  simulations, in: ASME International Mechanical Engineering Congress and
  Exposition, Vol. 57502, American Society of Mechanical Engineers, 2015, p.
  V08BT10A027.

\end{thebibliography}

%%%%%%%%%%%%%%%%%%%%%%%%%%%%%%%%%%%%%%%%%%%
% END DOCUMENT
%%%%%%%%%%%%%%%%%%%%%%%%%%%%%%%%%%%%%%%%%%%

\end{document}